\documentclass[prd,twocolumn,nofootinbib,superscriptaddress]{revtex4-1}
\usepackage{amsmath}
\usepackage{breqn}
\usepackage{graphicx}

\newcommand{\Rs}{R_{\rm S}}

\newcommand{\beq}{\begin{equation}}
\newcommand{\eeq}{\end{equation}}

\begin{document}

\title{Event Horizon Telescope Observations as Probes for Quantum Structure
of Astrophysical Black Holes}

\author{Steven B.\ Giddings}

\affiliation{Department of Physics University of California Santa Barbara CA 93106}

\author{Dimitrios Psaltis}

\affiliation{Departments of Astronomy and Physics University of Arizona Tucson AZ 85721}

\date{\today}

\begin{abstract}
The need for a consistent quantum evolution for black holes has led to
proposals that their semiclassical description is modified not just
near the singularity, but at horizon or larger scales.  If such
modifications extend beyond the horizon, they influence regions
accessible to distant observeration.  Natural
candidates for these modifications behave like metric fluctuations,
with characteristic length and time scales set by the horizon radius.
We investigate the possibility of using the Event Horizon Telescope to
observe these effects, if they have a strength sufficient to make
quantum evolution consistent with unitarity, without introducing new scales.  We find that such quantum
fluctuations can introduce a strong time dependence for the shape and
size of the shadow that a black hole casts on its surrounding
emission. For the black hole in the center of the Milky Way, 
 detecting the rapid
time variability of its shadow will require non-imaging timing
techniques. However, for the much larger black hole in the center of
the M87 galaxy, a variable black-hole shadow, if present with these
parameters, would be readily observable in the individual snapshots
that will be obtained by the Event Horizon Telescope.
\end{abstract}


\maketitle

\section{Introduction}

The discovery of Hawking radiation~\cite{Hawk} yields a logical
contradiction when one tries to account for quantum information
absorbed by a black hole: this information can't escape, can't be
destroyed, and can't be preserved after the black hole evaporates.
(For reviews, see~\cite{Preskill,Page,Trieste,Mathur,PhT,Harlow}.)
This situation appears to represent a fundamental conflict between the
principles underpinning local quantum field theory: the principles of
quantum mechanics, relativity, and locality.  Therefore, while it has
long been believed that the vicinity of the horizon is well-described
by classical general relativity, since curvatures are expected to be
small there, many theorists who study quantum evolution of black holes
have now concluded that there must be modifications to their
description via local quantum field theory, and that in order
to resolve the conflict, these modifications must extend {\it at least  to
  horizon scales}.

A variety of proposals have been made for such new physics.  One
possibility is new {\it hard} structure outside of, and replacing, the
horizon, {\it e.g.\/}, if a new kind of massive black hole
remnant~\cite{BHMR} forms, placing the information outside the horizon,
while the black hole is still large.  Variants of this basic scenario
include that of fuzzballs~\cite{Mathurfuzz}, where higher-dimensional
string theory configurations are proposed to replace the smooth
four-dimensional geometry outside the horizon, or that of
gravastars~\cite{gravastar}.  More recently, it has  similarly been
proposed~\cite{AMPS} that semiclassical geometry breaks down within a
Planck distance $l_{\rm Pl}$ of the horizon, forming a ``firewall"
that destroys all infalling matter.  So far, however, there is no
concrete theoretical description of any of these scenarios for
Schwarzschild or Kerr black holes in terms of more fundamental theory.
(Other scenarios propose even more radical modification of the
structure of spacetime, extending to greater distances from a black
hole~\cite{EREPR}.)

Such a hard  (high momentum) quantum structure for black holes represents
a major departure from the familiar spacetime description of black
holes and, in particular, greatly alters the description of the
infalling observer.  Moreover, any theory that accounts for it will
apparently have to explain very disparate scales.  To take the extreme
example, the firewall scenario assumes that new physics alters the
classical geometry of a black hole not just at the center of the black
hole but all the way to the horizon, {\it but} the modification
extends at most a distance $\sim l_{\rm Pl}$ (or some other
microscopic cutoff scale) from the horizon.  This appears to require
an extreme fine tuning for a large black hole.

Indeed, to begin to quantify differences in proposals, one may define
a scale $R_a$ which characterizes how far outside the would-be horizon
the semiclassical geometry is significantly modified; for the firewall
proposal, one thus has $R_a\sim l_{\rm Pl} \ll \Rs$, where $\Rs$ is the
Schwarzschild radius of the black hole.\footnote{For simplicity this paper treats non-rotating black holes as a test case, though our results are expected to generalize to the rotating case.}  Another characteristic
scale~\cite{SGwindows} is the scale $L$ on which such departures from
the semiclassical geometry vary.  Clearly $L\leq R_a$ -- one could
have a ``quantum atmosphere" extending to $R_a$, but with ``harder"
(higher momentum) variation on shorter scales.  For example, in the
fuzzball context~\cite{Mathurfuzz}, it has been suggested~\cite{warner}
that $L$ could be the size of the extra dimensions of string theory,
{\em i.e.\/}, microscopic, but there is no clear prediction for $R_a$, which
could be much larger.

If we want to avoid such a violent breakdown of the spacetime, {\it e.g.}, as
seen by the infalling observer, and the fine tuning we have described, 
the scales $L$ and $R_a$ should become
large for large black holes.  For example, they could scale as
$R_a\sim \Rs^p$, $L\sim \Rs^q$, for some $p,q>0$, so the momentum scales
of perturbations seen by infalling observers become soft for a large
enough black hole.  Refs.~\cite{SGmodels,BHQIUE,NVNL} proposed this
alternative: the new physics that produces quantum structure on black
holes is not finely tuned so that this structure ends abruptly at the
horizon, but can instead extend to a macroscopic distance outside, and
can be ``soft," or ``nonviolent."  Indeed, if the necessary new physics leads
to quantum modifications to the standard description over distances
$\sim \Rs$ (reaching from the center of the black hole to the horizon), the
simplest alternative is that this physics is characterized by this
single scale, so that $p=q=1$, and the quantum structure extends a
distance $\sim \Rs$ outside the horizon, rather than ending sharply
there.

Any such perturbations to the black-hole metric that extend to
  a distance $\sim \Rs$ can, of course, have observable implications for
  astrophysical phenomena that originate in the vicinities of 
  black-hole horizons~\cite{SGwindows,SGobs}. Traditionally, the
  profiles of iron fluorescence lines and the variability of the
  X-rays observed from accreting black holes have been used to probe
  such strong gravitational fields (see discussion in
  Ref.~\cite{Psaltis2008} and references therein).  Future missions,
  such as ESA's Athena, will have the capabilities to trace the time
  evolution of the fluorescence lines at the dynamical timescale of
  the innermost stable circular orbit around supermassive black
  holes~\cite{Nandra2013} and potentially observe such perturbations,
  if they extend to large enough radius.

In the near future two new approaches to probing the innermost regions
of black hole spacetimes will become available, with the potential of
offering probes that are clean of astrophysical complexities. The
first involves gravitational wave observations either of coalescing
black holes (such as the initial LIGO detection of the source
GW150914~\cite{Abbott2016}) or of extreme mass ratio
inspirals~\cite{eLisa2013}. The second involves obtaining images of
accreting black holes with horizon-scale resolution using the Event
Horizon Telescope~\cite{Doeleman2009a}. The prospect that
gravitational-wave observations could reveal quantum modifications to
black hole dynamics was preliminarily explored in
Ref.~\cite{Giddings2016}.

The goal of the current paper is to investigate the possible
signatures of soft quantum modifications to black hole metrics that
could be imprinted on Event Horizon Telescope
observations~\cite{SGwindows}, and thus the sensitivity of these
observations to such effects.  The benefit of using Event Horizon
  Telescope observations is that they are probing the stationary
  spacetimes of supermassive black holes, long after the decay of the classical 
  ringdown modes that were excited during their formation.  For the inferred accretion rates of the primary Event
  Horizon Telescope targets, the amount of mass in the accretion flow
  is a negligible fraction ($\sim 10^{-9}$) of the mass of the black
  hole and is not expected to excite any classical spacetime
  oscillations with amplitudes comparable to those we will be studying
  here.

In \S2, we discuss in more detail the origin and scale of
perturbations that might be relevant to large astrophysical black
holes.  Quantifying their effects on the predicted black hole images
requires performing ray tracing calculations from the image plane of a
distant observer down to the horizon of the black hole. For these
calculations to be possible, the perturbations to the metric that we
consider need to preserve its signature and avoid introducing
pathologies outside the horizon (see discussion in
Ref.~\cite{Johannsen2013a}). In \S3, we use the formalism of Regge \&
Wheeler~\cite{Regge1957} to write a general form of metric
perturbations, modulo gauge transformations, and to impose these
constraints.

Several earlier studies have explored the effect of
  stationary modifications to black-hole metrics on the images that
  will be generated with the Event Horizon
  Telescope~\cite{Johannsen2010a,Bambi2010, Abdujabbarov2012,
    Amarilla2013, Vincent2016}. However, introducing time-dependent,
  non-axisymmetric perturbations to a spacetime, such as the ones we
  discuss here, removes all Killing symmetries of the Kerr metric,
  which have been crucial in improving the performance of most ray
  tracing algorithms available today. In \S4, we discuss our
modifications to our existing algorithm in order to perform ray
tracing calculations in a time-dependent, non axisymmetric spacetime.

In \S4, we also perform a number of exploratory calculations for
different monochromatic perturbations of the black-hole metric and
identify the range of parameters that introduce observable effects on
the predicted images. In \S5, we use simple models for the plasma in
the accretion flow around a black hole in order to make concrete
predictions of the effects of strong, soft quantum perturbations on the image of
accreting black holes. Finally, in \S6, we discuss  implications of
our results for Event Horizon Telescope observations of the two
primary targets, the black hole in the center of the Milky Way and the
one in the center of the M87 galaxy.

\section{Quantum Structure for Astrophysical Black Holes}

The ultimate explanation of the quantum structure of a black hole may
lie in a fully quantum fundamental description of spacetime.  However,
we can ask what is a ``minimal" departure from the usual description
of black holes by semiclassical spacetime and local quantum field
theory that is sufficient to transfer quantum information from the
black hole state to the outgoing radiation, as is apparently necessary
to recover unitary quantum-mechanical evolution.\footnote{The needed
  information transfer can be sharply characterized in terms of
  transfer of entanglement~\cite{HaPr,GiShone,Sussxfer}.}  One expects
that such a minimal departure could be parameterized in terms of new
couplings between the ``internal" state of the black hole and the
quantum fields {\it near} the black hole -- where a quantum field
theory description is expected to be approximately valid -- and that
these couplings transfer the required
information~\cite{SGEFT,GiShtwo}.

A lower bound on the strength of such couplings is thus set by the
requirement that they should transfer of order one qubit of
information per time $\Rs$.  This is the necessary transfer rate from
a black hole once it has reached the midpoint of its
evaporation~\cite{Pageone,Pagetwo}, in order to cause a decrease of
its von Neumann entropy $S_{\rm vN}$ to zero, rather than the increase
to $S_{\rm vN}\sim \Rs^2$ predicted by Hawking.  In principle, a wide
variety of such couplings could accomplish this, {\it e.g.}, couplings
that just produce photon, or graviton, emission.

However, much theoretical work on black holes has been guided by the
beautiful story of black hole thermodynamics, in which the black hole
entropy $S_{\rm BH}$ is proportional to its horizon area, as proposed
by Bekenstein and Hawking.  If this story is to be preserved, at least
approximately, then this imposes a strong constraint on any new
dynamics of black holes~\cite{BHQIUE},\cite{GiShone},\cite{AMPS},\cite{NVNL},\cite{SGEFT},\cite{SGstat}.
For
example, if the information-transferring couplings produce a flux of
gravitons from the black hole, in addition to that from the Hawking
radiation, then one expects that, if the black hole is brought into
thermal equilibrium, it will no longer equilibrate at the Hawking
temperature and thus will have entropy different from $S_{BH}$.
Moreover, a coupling {\it just} to gravitons would appear to violate
detailed balance and prevent equilibration. In order to preserve an
approximation of the standard thermodynamic description, we therefore
expect that we should find universal couplings to the fields that can
radiate from the black hole, and that these should closely match the Hawking radiation.

In fact, there is a second argument for such
universality~\cite{NVNL,GiShtwo,SGwindows}, based on gedanken experiments\cite{AMPS}
involving black hole mining~\cite{FrFu,Frol,LaMa,Brown}.  Here, a
cosmic string, or other apparatus, is introduced into the black hole
atmosphere to increase its emission rate.  To preserve unitarity, the
information transfer out of the black hole must increase
commensurately.  This suggests the necessity of universal couplings to any
possible kind of mining apparatus.  

Such universal couplings are most
naturally achieved via a coupling between the black hole state and the
stress tensor, {\it e.g.}, via a term in the action~\cite{SGEFT,SGmod}
\beq
\label{effint}
\Delta S = \int d^4x \sqrt{-g} H^{\mu\nu}(x) T_{\mu\nu}\ .
\eeq
Here $H^{\mu\nu}(x)$ is an {\it operator} that has nontrivial matrix
elements between black hole quantum states and $T_{\mu\nu}$ is the
stress tensor of all quantum fields (including the graviton).  The
integral extends over the support of $H^{\mu\nu}(x)$, which we take to
typically extend to radii $\sim \Rs+R_a$ from the black hole center.
The universality of \eqref{effint} nicely mirrors the universal nature
of gravity.

In order to produce an average emission spectrum without large
departures from the thermal Hawking spectrum, typical matrix elements
of $H^{\mu\nu}(x)$ should have time variation on scales $\Delta t\sim
\Rs$; hard structure can be avoided if the spatial variation is on
scales $|\Delta {\vec x}|\sim L$, as described above.  Then, the
typical magnitude of the elements of $H^{\mu\nu}(x)$ is determined by the
condition that they transfer information (or entanglement) at the rate
stated above, $\sim 1$ qubit per time $\Rs$.  In fact, since
$H^{\mu\nu}(x)$ is dimensionless, when $R_a\sim L\sim \Rs$ the magnitude
of $H$ most simply follows from dimensional analysis: we
expect~\cite{SGmod} a typical size $\langle H^{\mu\nu}\rangle \sim 1$
over the nontrivial range $\Delta x\sim R_a\sim \Rs$, as is seen in more
detailed analysis.  Put more simply, this condition arises from the need for an 
${\cal O}(1)$ modification to the Hawking radiation, to restore unitary evolution.

If in eq.~\eqref{effint} we replace the operator $H$ by its average
$\langle H^{\mu\nu}\rangle$ over black hole quantum states, we see
that these interactions change the action in the same way 
as if we were to consider a classical perturbation of the metric,
$g_{\mu\nu}\rightarrow g_{\mu\nu}+h_{\mu\nu}$, given by (to linear
order) 
\beq
h_{\mu\nu}(x) = -2 g_{\mu\kappa} g_{\nu\lambda} \langle H^{\kappa\lambda}(x)\rangle\ .
\eeq
And if, as the simplest proposal outlined above yields, $\langle
H\rangle \sim 1$, the metric perturbations are not parametrically
small: the quantum structure is soft, but {\it strong}.  That is, the
metric perturbations that suffice to accomplish the needed information
transfer are significant -- though curvatures near the horizon remain
small~\cite{SGmod}.

To reiterate, the reasoning explained above sets the range $R_a$
outside the horizon over which this {\it effective} metric
perturbation is nonvanishing, the scale $L$ of its variation, and its
magnitude.  In the absence of a complete theory of quantum gravity,
which might allow us to derive the matrix elements of $H^{\mu\nu}$, at
least approximately, we can investigate the sensitivity of
astrophysical observations to this type of strong, soft effective metric
fluctuation by considering relatively general perturbations with these
characteristic scales.  While we specifically focus in this paper on
the case $R_a\sim L\sim \Rs$, this obviously could be generalized to
describe scenarios with different characteristic scale sizes; one can
likewise generalize the couplings \eqref{effint} to give an effective
description of other proposals for black hole quantum
interactions~\cite{SGEFT,GiShtwo} -- albeit ones that may do more
damage to black hole thermodynamics. In order to stay close to the
thermodynamic description, We will specifically focus on perturbations
near the thermal frequency, $\omega_T=1/(8\pi M)$ for a Schwarzschild
black hole (unless explicitly noted otherwise).  This also avoids introducing a new scale in the physics.
As we note below, it is also true that higher (or lower) frequency perturbations have smaller effect
on black hole images.

\section{Perturbed metrics for black holes}

In this initial study of possible observational effects of quantum
structure of black holes, we will focus our attention on non-spinning
black holes.  In order to minimize artifacts due to coordinate
singularities, we will work in a coordinate system that is regular at
the future horizon. One such suitable choice is that of ingoing
Eddington-Finkelstein coordinates $(v,r,\theta,\phi)$, where the
metric $g_{\mu\nu}$ has a form with line element 
\begin{equation}
  ds^2=-\left(1-\frac{\Rs}{r}\right)dv^2+2 dv dr +r^2 (d\theta^2+ \sin^2\theta
  d\phi^2)\;.
\end{equation}
Unless otherwise noted, we set hereafter $G=c=M=1$, where $G$,
  $c$, and $M$ are the gravitational constant, the speed of light, and
  the mass of the black hole, respectively.

As discussed in \S2, we will consider linearized perturbations to this
metric, which we will denote by ${h}_{\mu\nu}$, such that the perturbed
metric becomes
\begin{equation}
  \tilde{g}_{\mu\nu}=g_{\mu\nu}+h_{\mu\nu}\;.
  \end{equation}
If these are treated as classical perturbations, not all such
perturbations are physical, since some can be removed by a coordinate
transformation.  Such a transformation, with parameter $\xi^\mu$,
produces a linearized change
\beq
\delta h_{\mu\nu} = \nabla_\mu \xi_\nu +  \nabla_\nu \xi_\mu\ .
\eeq
The analysis of Regge and Wheeler~\cite{Regge1957} shows that using
this freedom, a general perturbation can be reduced to a sum of even and odd
perturbations with specific forms.  

The even perturbations
take the form
\begin{equation}
  {h}_{\mu\nu,{\rm even}}=\left(\begin{array}{cccc}
     h_{vv} & h_{rv} & 0 & 0\\
    h_{rv} & h_{rr} & 0 & 0\\
    0 & 0 & r^2 \gamma_{\theta\theta} & 0 \\
    0 & 0 & 0 & r^2\sin^2\theta \gamma_{\theta\theta}
  \end{array}
  \right)
\end{equation}
and the odd perturbations the form
\begin{equation}
  {h}_{\mu\nu,{\rm odd}}=\left(\begin{array}{cccc}
    0 & 0 & h_{v\theta} & h_{v\phi}\\
    0 & 0 & h_{r\theta} & h_{r\phi}\\
    h_{v\theta} & h_{r\theta} & 0 & 0 \\
    h_{v\phi} & h_{r\phi} & 0 & 0
  \end{array}
  \right)\;.
\end{equation}
The quantities $h_{vv}$, $h_{rv}$, $h_{rr}$, and $\gamma_{\theta\theta}$ can be arbitrary functions of $(v,r,\theta,\phi)$. Thus they can be expanded in scalar spherical harmonics $Y_{lm}$,
\beq
h_{\alpha\beta}=\sum_{lm} f^{lm}_{\alpha\beta}(v,r) Y_{lm}\ ,
\eeq
where $\alpha$ and $\beta$ range over values $v$ and $r$, and
\beq
\gamma_{\theta\theta}=\sum_{lm} f^{lm}_{\theta\theta}(v,r) Y_{lm}\ \ .
\eeq
The quantities $h_{va}$ and $h_{ra}$, with $a=\theta,\phi$, can be expanded in terms of vector spherical harmonics,
\beq
h_{va}=\sum_{lm} f_{v}^{lm}(v,r) Y^a_{lm}\quad ,\quad h_{ra}=\sum_{lm} f_{r}^{lm}(v,r) Y^a_{lm}\ ,
\eeq
given by
\beq
Y^\theta_{lm} = -\frac{1}{\sin \theta} \partial_\phi Y_{lm}\quad ,\quad Y^\phi_{lm} = {\sin \theta} \partial_\theta Y_{lm}\ .
\eeq

The $v$ and $r$ dependence of these perturbations can then be described by
expanding the $f^{lm}$ elements in terms of a convenient basis of functions
of these coordinates, {\it e.g.}, plane waves in $v$ and $r$.  However,
as we have discussed above, we consider the (most conservative)
case~\cite{SGmodels,BHQIUE,NVNL} where such perturbations are well-localized within
a range of size $R_G$ around the horizon radius $\Rs$, where $R_G\sim
\Rs$.  Thus, we specifically consider (sums of) perturbations of the
form
\begin{eqnarray}
f^{lm}_A(v,r) &=& \int d\omega dk\, \epsilon_A^{lm}(\omega, k)  \exp\left[-\frac{(r-\Rs)^2}{2R^2_{\rm G}}\right]\nonumber \\
&&\quad exp\left[-i \left(\omega v- kr\right)\right] \;
\end{eqnarray}
parameterized by the functions $\epsilon_A^{lm}(\omega, k)$, where the
index $A$ takes on the values $vv, rr, rv, \theta\theta,v,r$.  These
functions of Fourier variables characterize the spectrum of
perturbations that we consider.  We may alternately chose to superpose
such plane waves so that they localize about a specific time $v$.

In general, such perturbations can lead to
pathologies~\cite{Johannsen2013a}.  Specifically, avoiding change in
signature of the metric restricts the size of the perturbations.
Considering for example the even perturbations, avoiding signature
change clearly requires $1+\gamma_{\theta\theta}>0$, and the condition
\begin{equation}
  {\rm Det}\left\vert\begin{array}{cc}
  -1+\frac{\Rs}{r}+h_{vv} & 1+h_{rv} \\
  1+h_{rv} & h_{rr}
  \end{array}\right\vert<0\;,
\end{equation}
which translates to
\begin{equation}
  h_{rr}\left[1-\frac{\Rs}{r}-h_{vv}\right]+
  \left(1+h_{rv}\right)^2>0\;.
\end{equation}
Similar conditions can be derived for the odd perturbations and more
generally for combined perturbations.  In the calculations
  presented in this paper, we ensure that the perturbations we
  consider do not violate these no-signature change conditions, using
  the above analytic considerations for individual modes or equivalent
  numerical conditions for simulations with superpositions of modes.
  For the latter case, we ensure numerically that the signature of the
  metric does not change throughout the simulation outside an excision
  region (set at $r=2.1$), and discontinue the integration of the
  geodesics if such a change occurs inside this excision region.

\begin{figure*}
  \includegraphics[scale=0.4, bb=3 7 510 461]{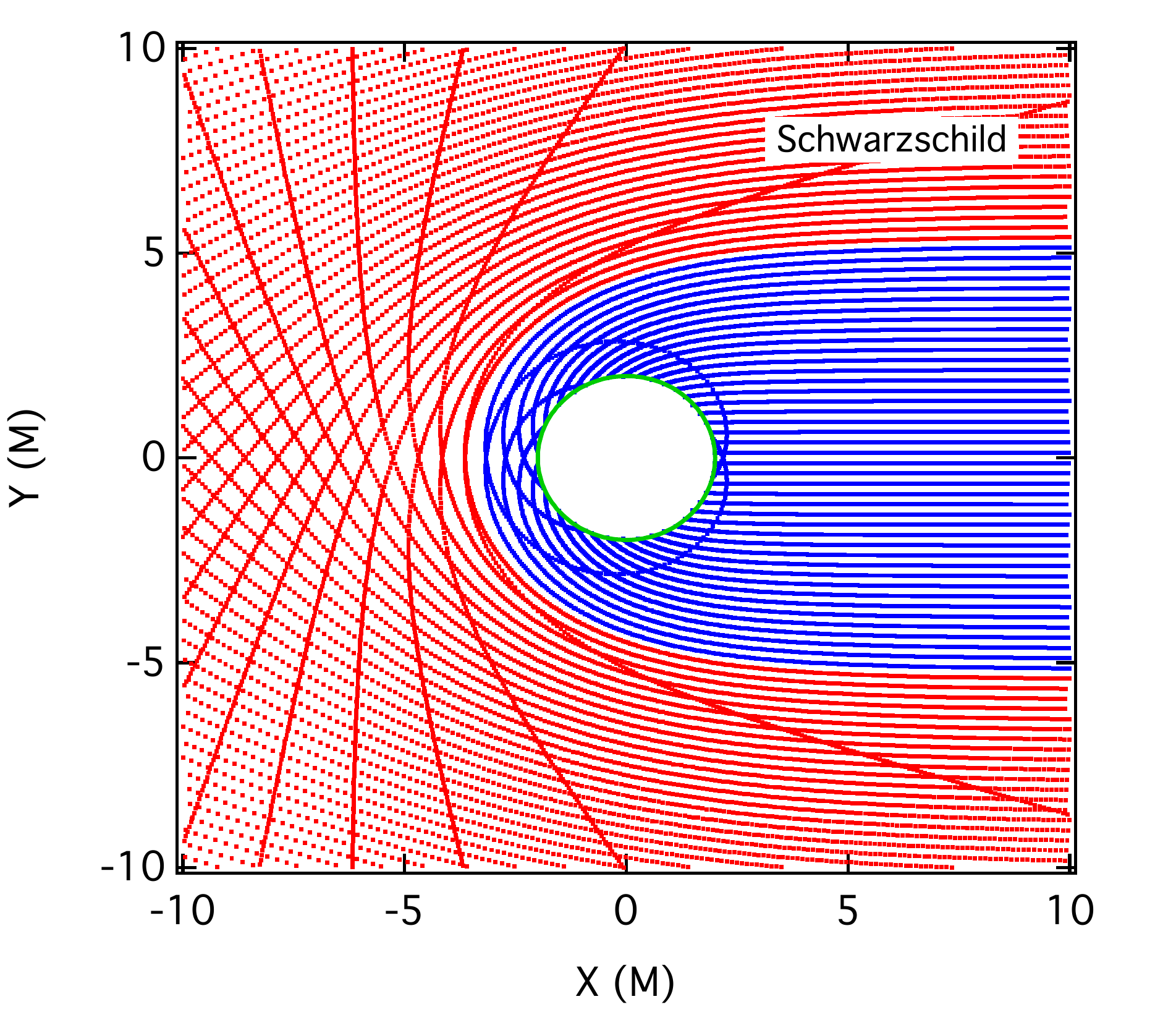}
  \includegraphics[scale=0.4, bb=3 7 510 461]{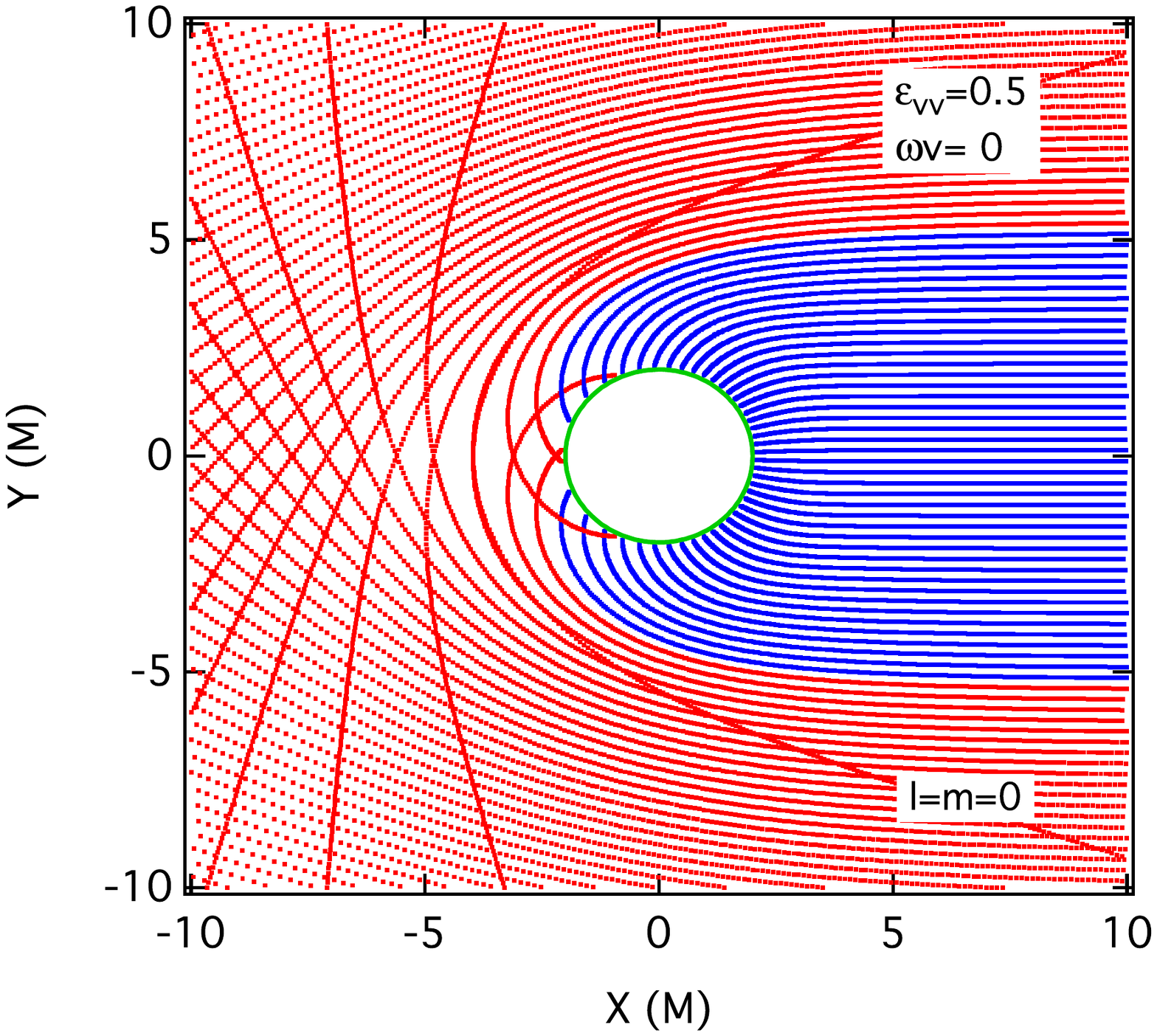}
  \includegraphics[scale=0.4, bb=3 7 510 461]{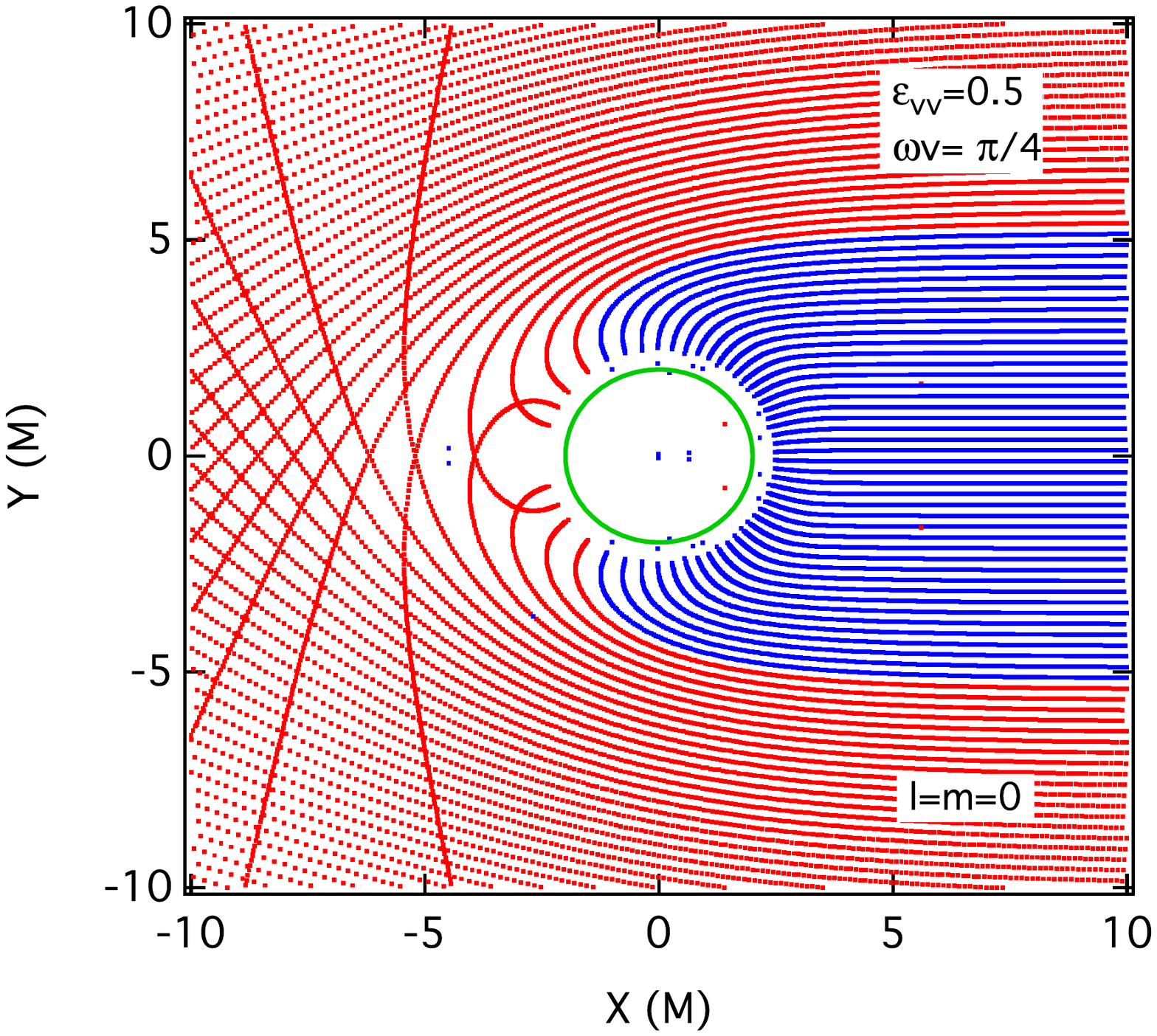}
  \includegraphics[scale=0.4, bb=3 7 510 461]{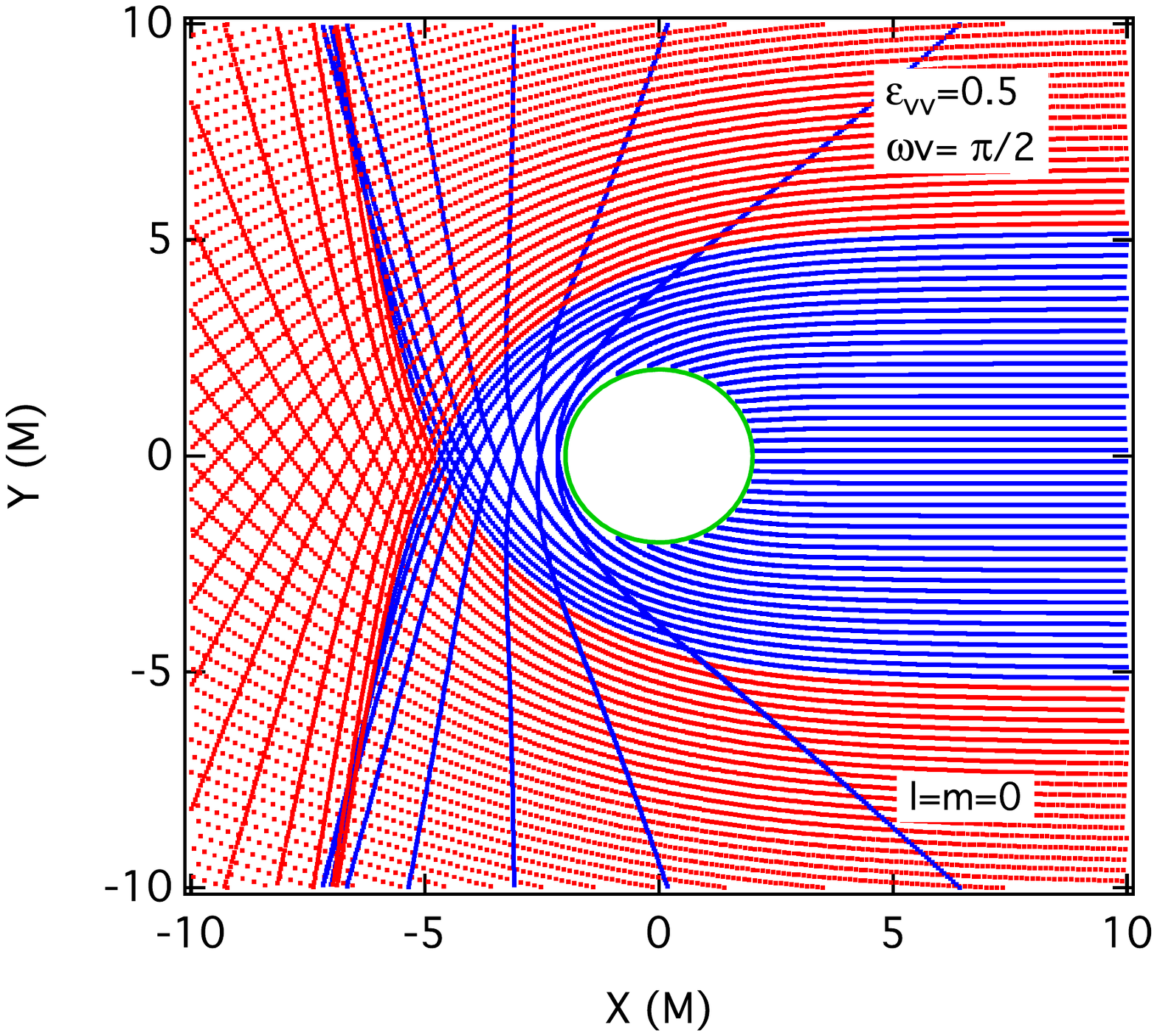}
  \caption{\footnotesize The trajectories of photons approaching a non-spinning
    black hole in plane parallel rays from the far infinity (from the
    right side of the figure). The upper left panel shows the
    trajectories around an unperturbed Schwarzschild black hole. The
    other panels correspond to different phases of a black hole with a
    single, spherically symmetric metric perturbation (see text for
    details). In all panels, photon trajectories are colored blue if
    (in the absence of any perturbations) they would have crossed the
    event horizon and are colored red, if they would have escaped to
    infinity. As expected, metric perturbations cause some of the blue
    trajectories to escape to inifnity and some of the red
    trajectories to cross the event horizon, at different phases of
    the perturbation.}
\label{fig:trajectories}
\end{figure*}

\begin{figure*}
  \includegraphics[scale=0.4, bb=3 7 510 461]{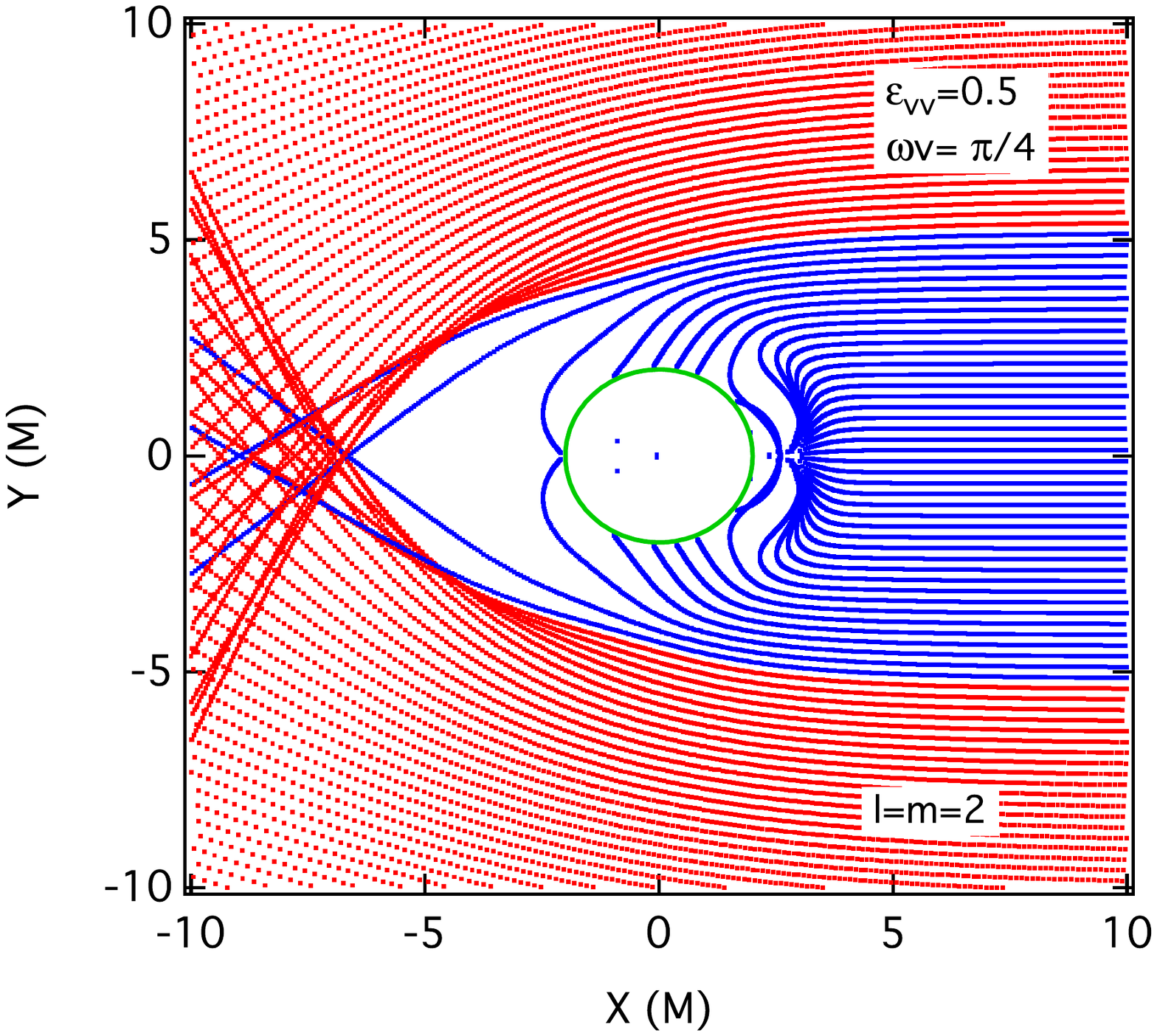}
  \includegraphics[scale=0.4, bb=3 7 510 461]{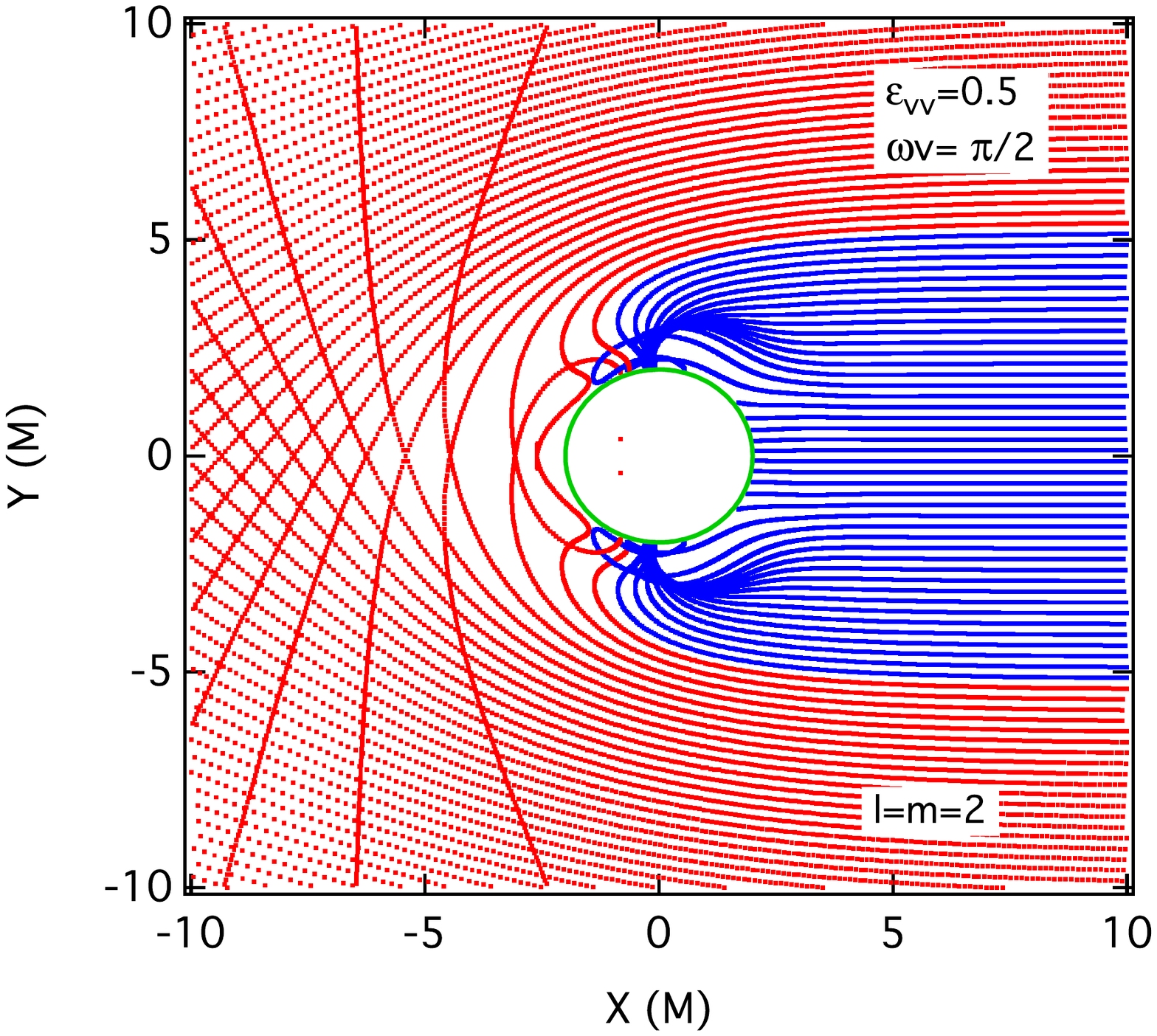}
  \caption{\footnotesize Same as in Figure~\ref{fig:trajectories} but
    for two phases of an $l=m=2$ perturbation mode. All other parameters
    remain the same.
  \label{fig:trajectories_harm}}
\end{figure*}

\begin{figure*}
  \includegraphics[scale=0.4, bb=3 7 510 461]{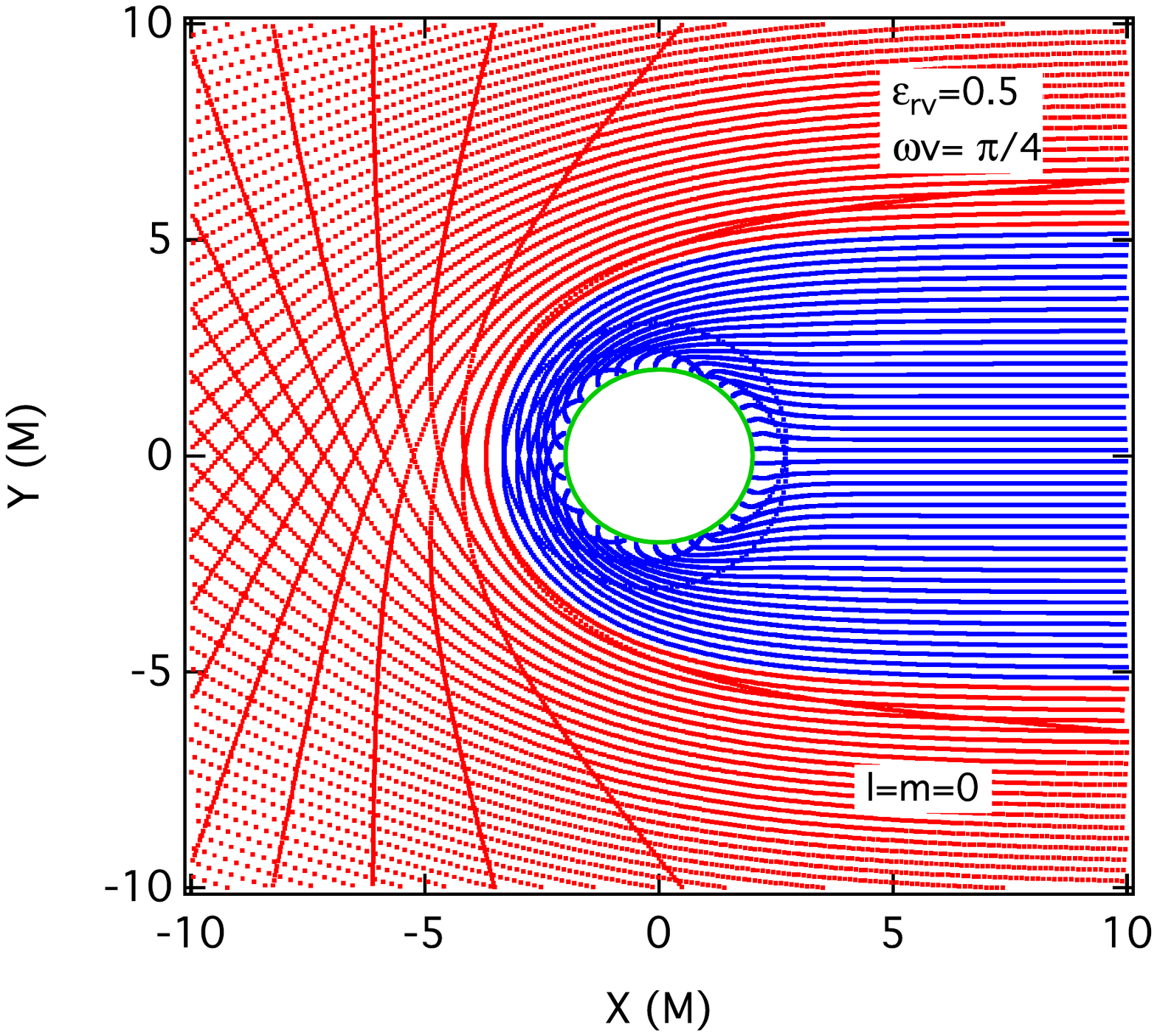}
  \includegraphics[scale=0.4, bb=3 7 510 461]{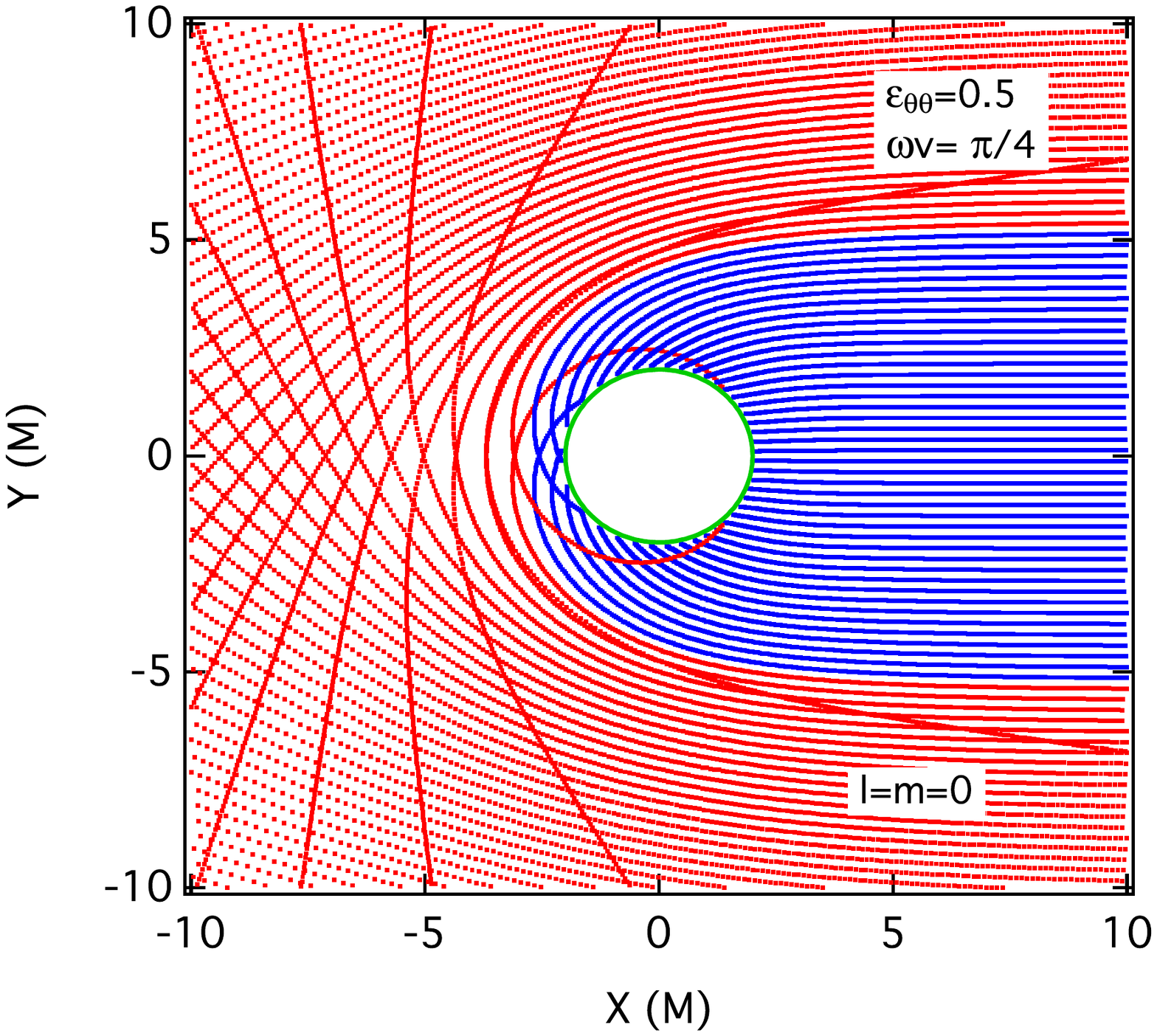}
  \caption{\footnotesize Same as in Figure~\ref{fig:trajectories} but
    for perturbing only the {\em (left)\/} $rv-$ and {\em (right)\/}
    $\theta\theta-$ components of the metric. All other parameters
    remain the same. The effects of perturbing different components of
    the metric are all qualitatively similar to those shown in
    Figure~\ref{fig:trajectories} for the $vv-$component.}
  \label{fig:trajectories_other}
\end{figure*}

\section{The Effect of Metric Perturbations on Photon Propagation}

In order to calculate the effect of metric perturbations on the
propagation of photons in the vicinity of the black-hole horizon, we
use the ray tracing algorithm described in Ref.~\cite{Psaltis2012a},
but with two important modifications. First, we rewrote the relevant
expressions for the metric and for the Christoffel symbols in terms of
the ingoing Eddington-Finkelstein coordinates we use here.  Second,
because the introduction of time-dependent, non-axisymmetric
perturbations removes all symmetries from the metric, we cannot use
any of the Killing equations to integrate the null geodesics. Instead,
we integrate directly the four second-order geodesic equations for the
coordinates $u, r, \theta$, and $\phi$, with the Christoffel symbols
 evaluated analytically to linear order in the metric perturbations.

The ray tracing algorithm considers a Cartesian grid $(\alpha_0,
\beta_0)$ on a virtual screen of an observer at infinity (the image
plane) and solves  using backwards evolution for the null
geodesics of photon trajectories that cross the image plane with
3-momenta that are perpendicular to the screen. Using these geodesics,
we define the shadow of the black hole as the locus of points on the
image plane at which the corresponding null geodesics cross the event
horizon of the black hole, when traced
backwards~\cite{Bardeen1973,Luminet1979}. When we calculate images
using the thermodynamic properties of plasma that fills the black hole
spacetime, we integrate the radiative transfer equation along each
null geodesic, neglecting the effects of scattering. For more details
of the ray tracing algorithm see Refs.~\cite{Psaltis2012a, Chan2013,
  Chan2015}

In order to understand the dependence of our results on the parameters
of the spectrum of perturbations, we first explore some simple cases
of monochromatic perturbations.  Figure~\ref{fig:trajectories} shows a
cross section of photon trajectories around a Schwarzschild black hole
and compares them to those in a perturbed spacetime, in which we
consider different phases of a single, spherically symmetric ($l=m=0$)
perturbation in the $vv-$component of the metric with an amplitude of
$\epsilon_{vv}^{00}=\epsilon_{vv}=0.5$, a radial extent of $R_{\rm
  G}=1$, a frequency of $\omega=\omega_{\rm T}=(8\pi)^{-1}$, and a
radial wavelength of $k_r=\omega_{\rm T}$. In all panels, the
trajectories are colored blue if, in the absence of any perturbations,
they would have crossed the event horizon and red, otherwise. Clearly,
the presence of perturbations affects the photon trajectories, causing
some of the photons that would have crossed the horizon to escape and
vice versa. This will significantly influence the shape and size of
the shadow cast by the black hole on the surrounding
emission. Moreover, the radial character of these perturbations causes
light rays with different impact parameters to periodically bundle up
and diverge, which will alter the brightness amplification introduced
by gravitational lensing and lead to bright structures in the
resulting images.

As Figures~\ref{fig:trajectories_harm} and
\ref{fig:trajectories_other} show, these qualitative properties of the
photon trajectories in perturbed spacetimes remain the same when we
introduce modes that are not spherically symmetric, or similar
perturbations in the other components of the metric. In the latter
case, we find that the amplitude of the effect is largest for
perturbations  to the $vv-$component of the metric. In fact,
perturbing the $rv-$component of the metric has the smallest effect on
the photon trajectories, at all phases of the oscillations. This is
expected since it is the $vv-$component of the metric that primarily
determines the magnitude of strong-field lensing experienced by the
photons in the vicinity of the black hole.

\begin{figure*}
  \includegraphics[scale=0.4, bb=3 7 510 461]{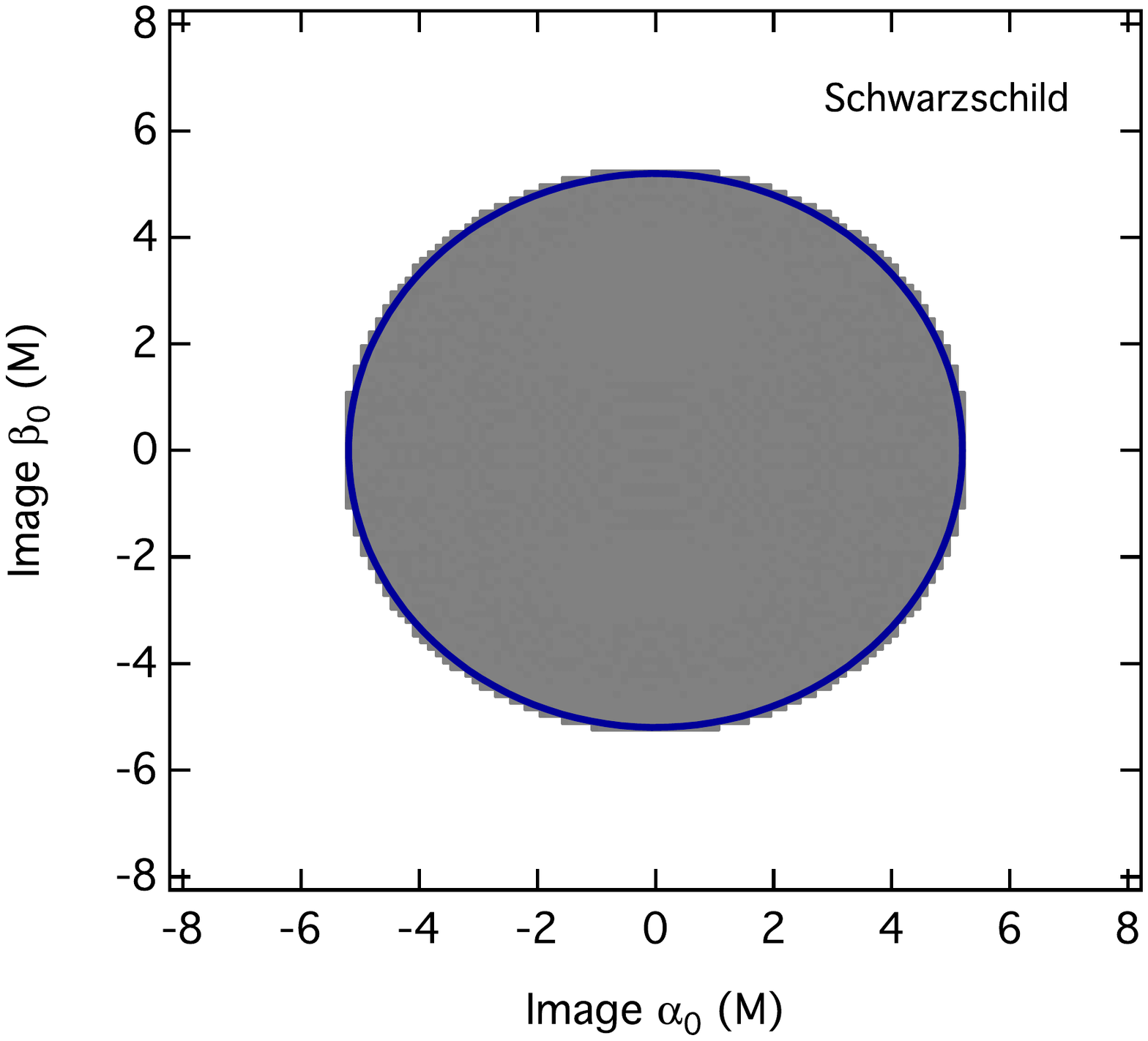}
  \includegraphics[scale=0.4, bb=3 7 510 461]{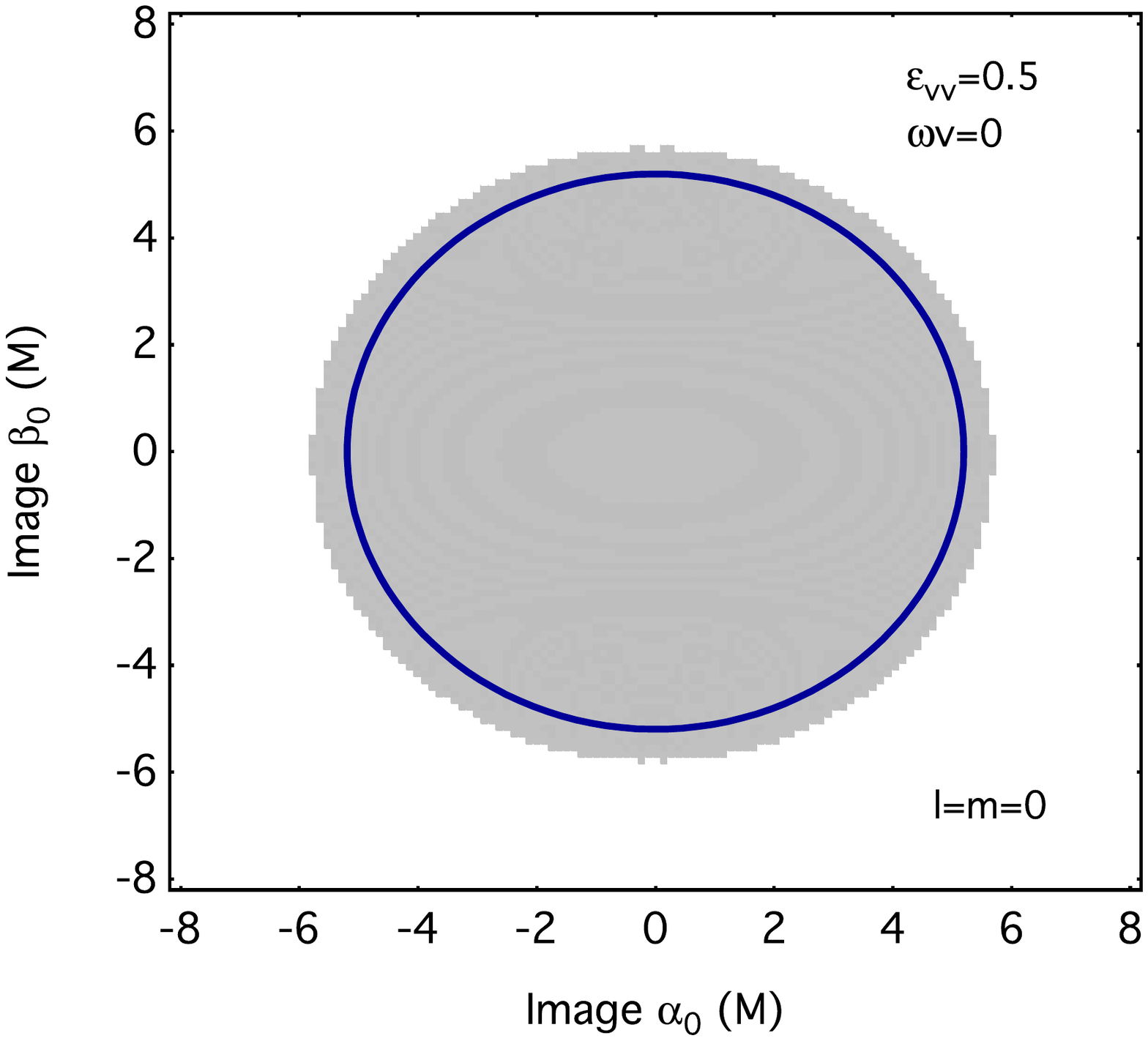}
  \includegraphics[scale=0.4, bb=3 7 510 461]{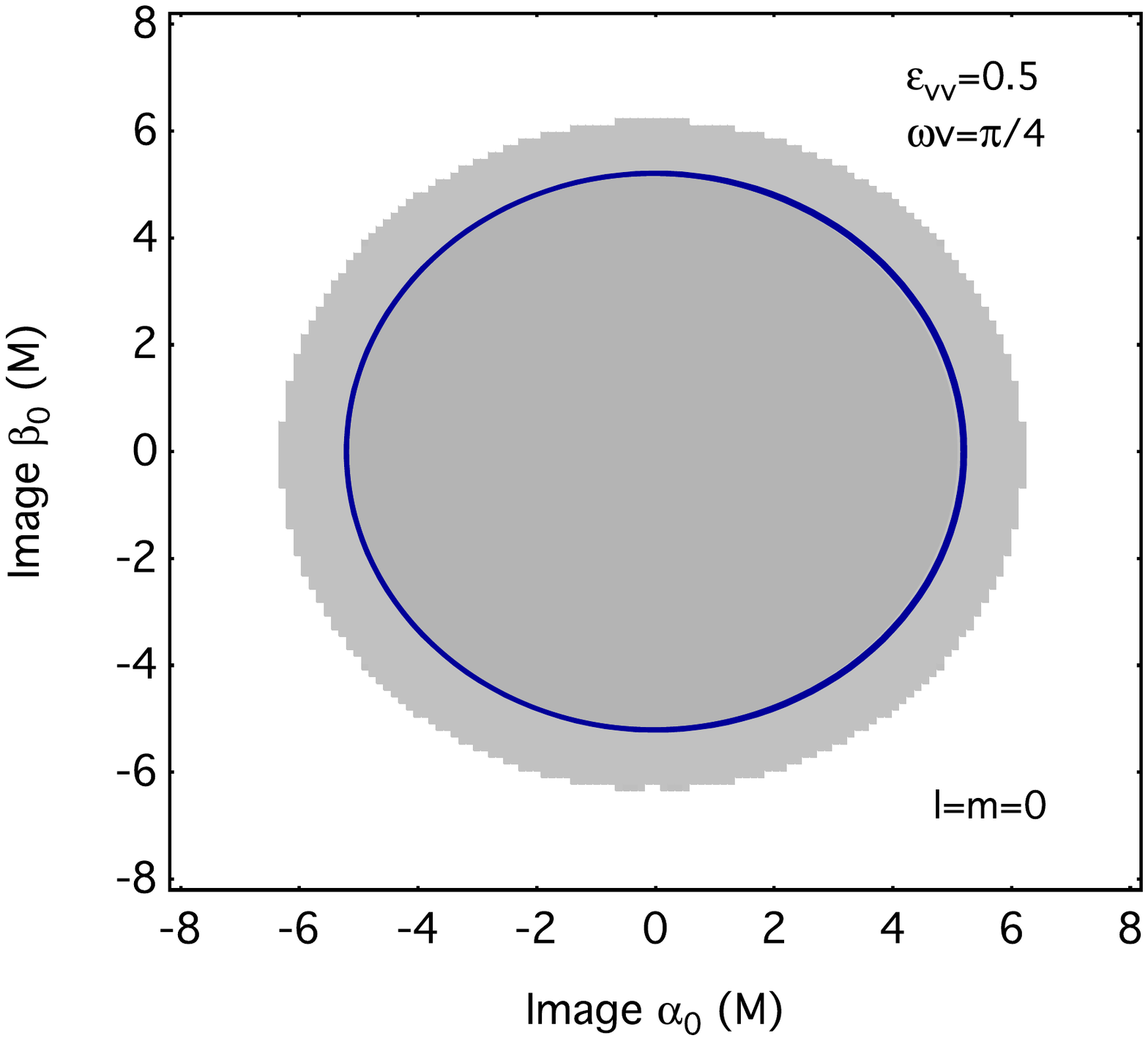}
  \includegraphics[scale=0.4, bb=3 7 510 461]{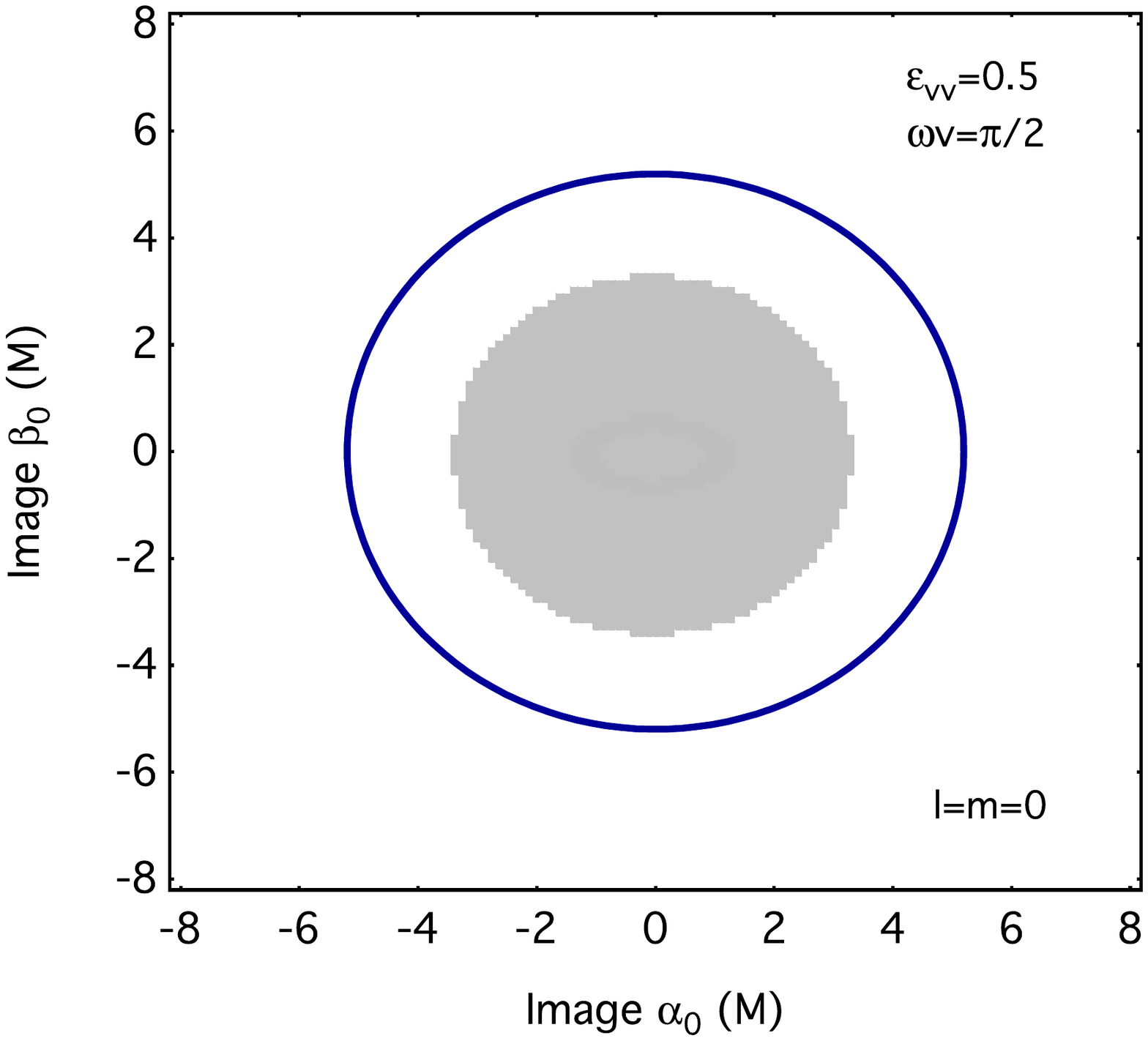}
  \caption{\footnotesize The shadows of the black holes that
    correspond to the four configurations shown in
    Figure~\ref{fig:trajectories}. In all panels, the blue circle
    shows the outline of the unperturbed Schwarzschild black hole.
    For the purturbed black hole, the size of the shadow is
    time-dependent and can be significantly different from the
    Schwarzschild case.}
\label{fig:shadows}
\end{figure*}

\begin{figure*}
  \includegraphics[scale=0.4, bb=3 7 510 461]{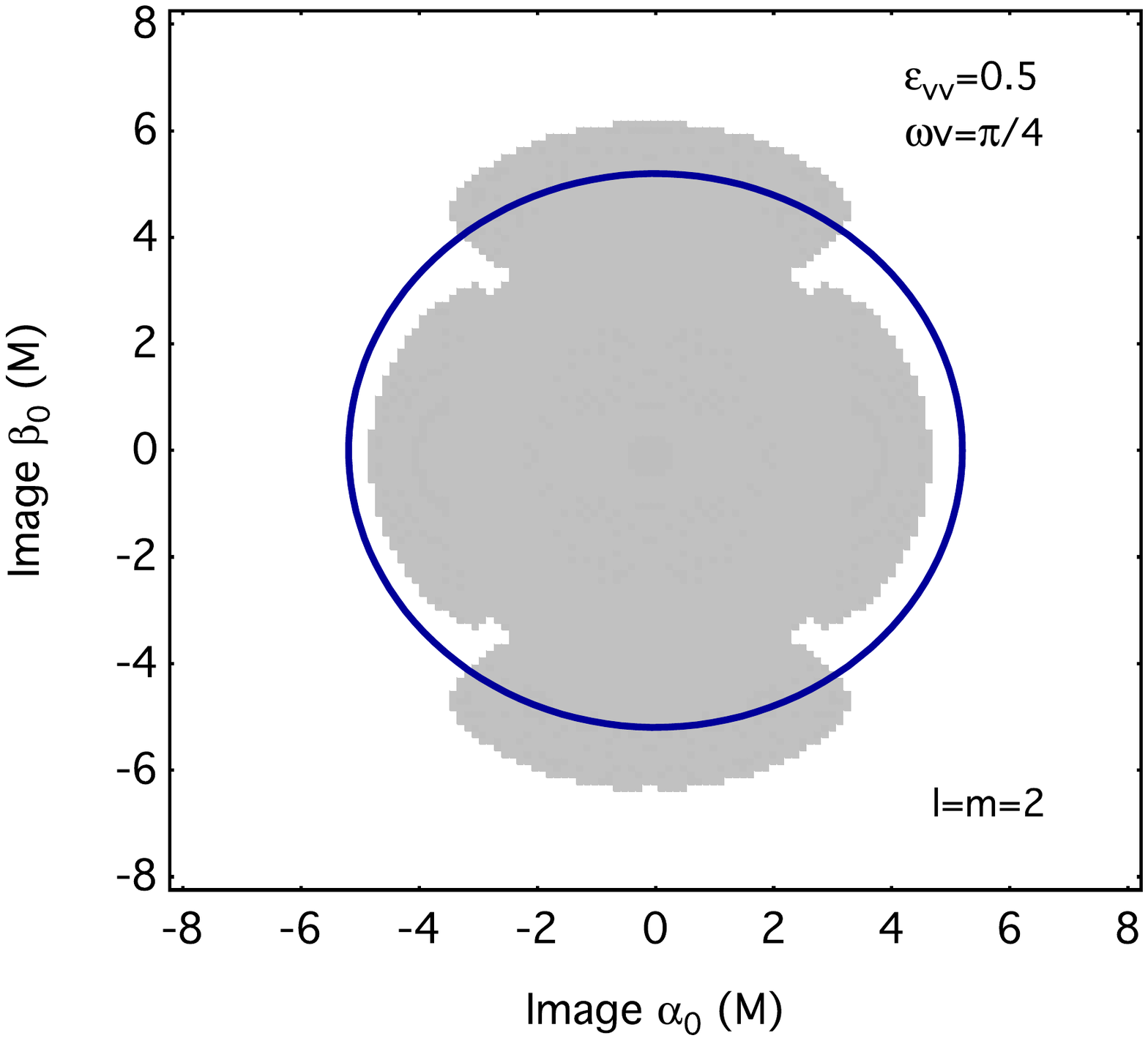}
  \includegraphics[scale=0.4, bb=3 7 510 461]{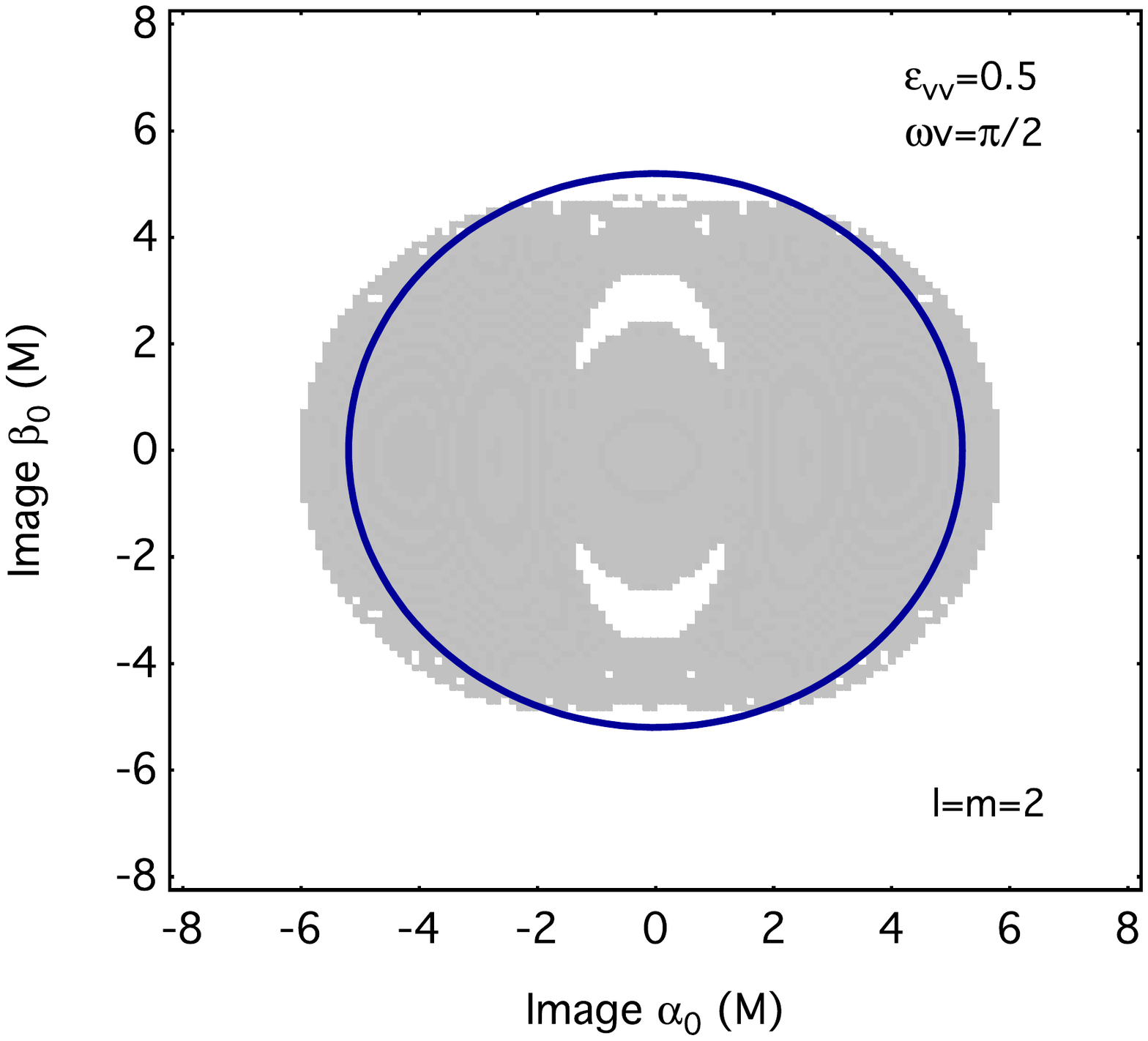}
  \caption{\footnotesize Same as in Figure~\ref{fig:shadows} but
    for two phases of an $l=m=2$ perturbation mode.
  \label{fig:shadows_harm}}
\end{figure*}

\begin{figure*}
  \includegraphics[scale=0.4, bb=3 7 510 461]{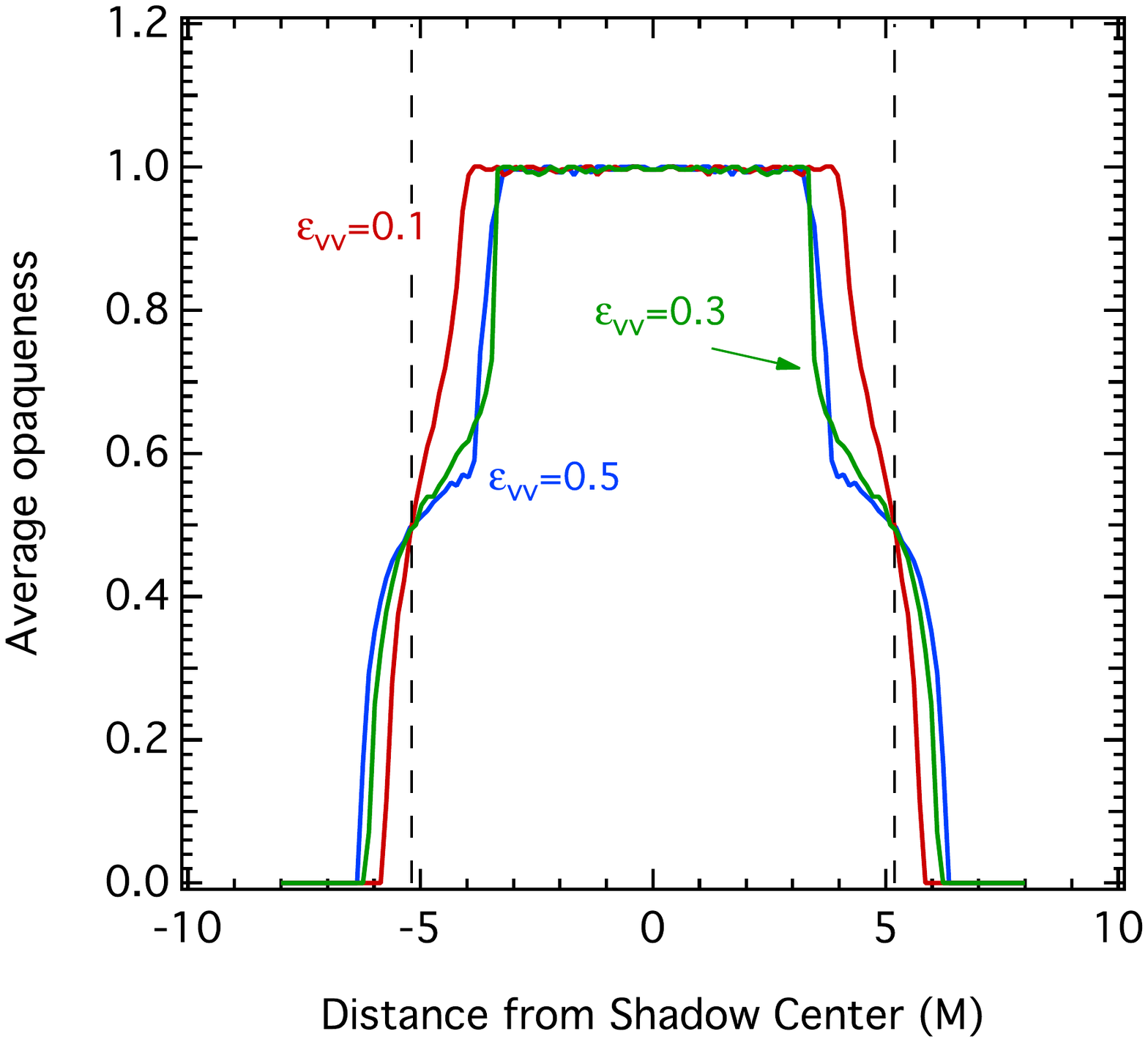}
  \includegraphics[scale=0.4, bb=3 7 510 461]{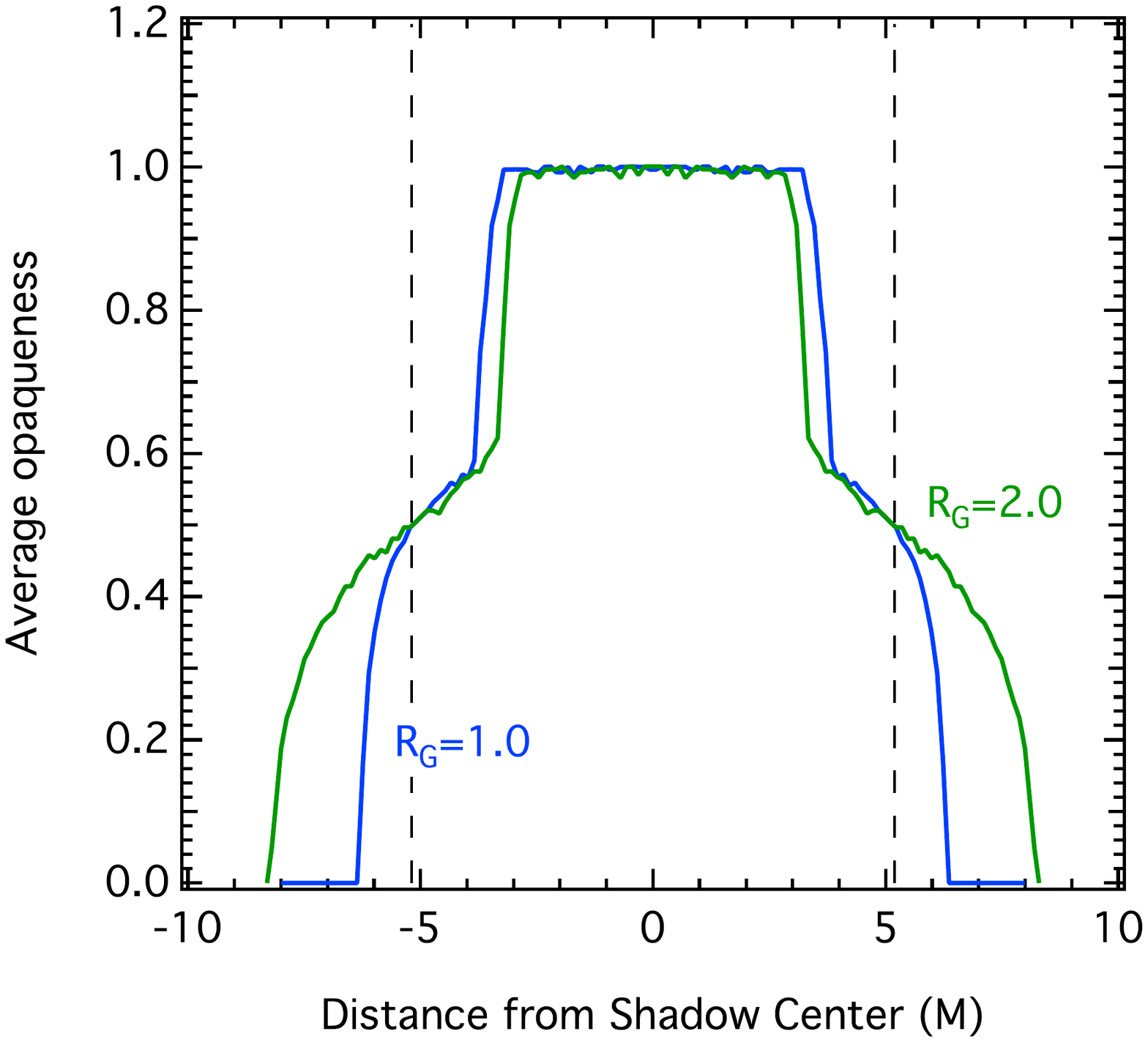}
  \includegraphics[scale=0.4, bb=3 7 510 461]{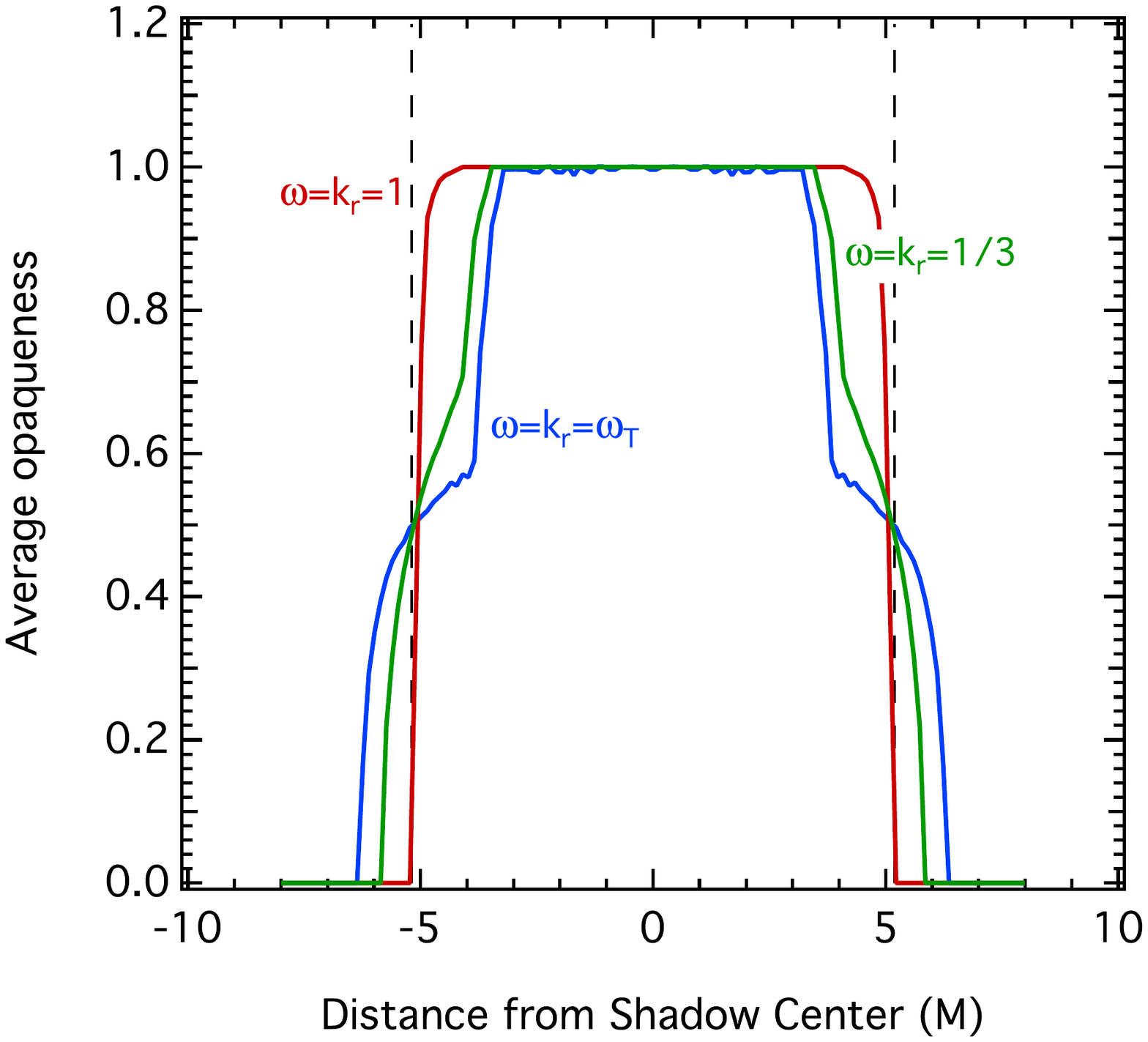}
  \includegraphics[scale=0.4, bb=3 7 510 461]{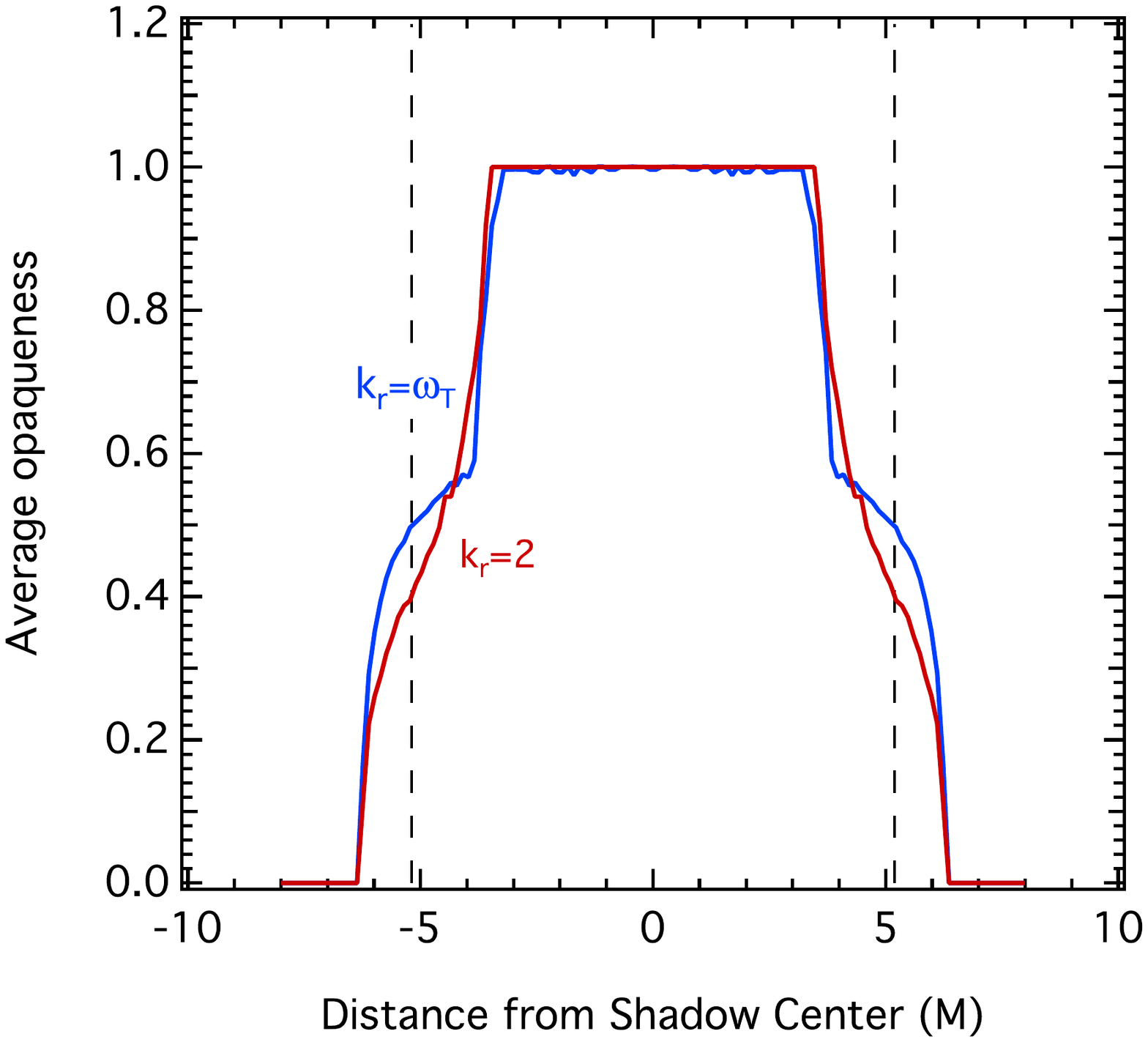}
  \caption{\footnotesize The dependence of the average opaqueness of
    the black-hole shadow (as defined in the text) on {\em (upper
      left)\/} the amplitude, {\em (upper right)\/} the range, {\em
      (lower left)\/} the frequency, and {\em (lower right)\/} the
    wavelength of the perturbations. In all panels, the vertical
    dashed line shows the location of the shadow for the unperturbed
    black hole and the values of the parameters not shown are the same
    as in Figure~\ref{fig:trajectories}. Introducing perturbations to
    the black-hole metric causes the black-hole shadow to loose its
    sharpness and become blurry. The effect is larger for
    perturbations that are strong, slow, and that extend to a large
    distance from the event horizon.}
\label{fig:opaq}
\end{figure*}

Figure~\ref{fig:shadows} shows the time dependence of the shadow cast
by the black hole on the surrounding emission for the different phases
of the spherically symmetric perturbation shown in
Figure~\ref{fig:trajectories}.  As expected, the size of the shadow is
a strong function of the perturbation phase; the shape of the shadow
remains circular because of the spherically symmetric character of the
perturbations considered here (see, however,
Figure~\ref{fig:shadows_harm} for a more general axisymmetric
case). The location and shape of the shadow is determined by the
properties of the circular photon orbit, which is located at $r=3$ in
the unperturbed spacetime. The circular photon orbit is unstable since
it corresponds to a local maximum in the effective potential for the
photons. As a result, even small perturbations in the metric can
significantly shift the location of the photon orbit and the effect
will only be amplified by the gravitational lensing that photons
experience as they travel from the vicinity of the black hole to the
observer at infinity.  In particular, note that we have set
$\epsilon_{vv}=0.5$ and the perturbation of the metric at the
location of the photon orbit has a magnitude of only
$\epsilon_{vv}\exp[-(3-2)^2/2]\simeq 0.3$. Even then, the net result
is a large periodic fluctuation of the diameter of the black hole
shadow that, in this case, ranges from $6M$ (lower right panel) to
$12M$ (lower left panel).

In order to explore the dependence of the average black-hole shadows
on the various parameters that describe the metric perturbations, we
define the average opaqueness of a given location on the observer's
image plane as the fraction of the perturbation period during which
this location is within the perimeter of the instantaneous shadow. For
the case of the spherically symmetric perturbations of
Figure~\ref{fig:trajectories}, the average opaqueness depends only on
the distance from the center of the black-hole shadow and is shown for
different values of the parameters of the perturbation in
Figure~\ref{fig:opaq}.  The radius of the photon orbit ($r=3$) in the unperturbed
metric and the light crossing time that corresponds to this length
($\delta v=3$) set the optimal lengths and frequencies for the
perturbations with maximal effects on the black-hole shadow. Indeed,
the changes in the black-hole shadow become appreciable for order
unity perturbations in the metric ($\epsilon_{vv}\gtrsim 0.1$), for
radial extent of order the horizon scale ($R_{\rm G}\gtrsim 1$), and
for relatively slow angular frequencies ($\omega \lesssim 1$).
Perturbations with smaller amplitudes that are confined too close to
the horizon do not affect the propagation of the photons in the
vicinity of the photon radius and hence have no appreciable effect on
the black-hole shadow. This is also true for perturbations that are
too fast compared to the light crossing time for the photon orbit,
since their effects average out during the passage of the photons
across the region that determines the black-hole shadow. Finally, as
expected, the radial wavelength of the perturbations alters the radial
dependence of the opaqueness as it determines the details of the
metric perturbations within the envelope described by the range
$R_{\rm G}$.

\begin{figure*}
  \includegraphics[scale=0.4, bb=3 7 510 461]{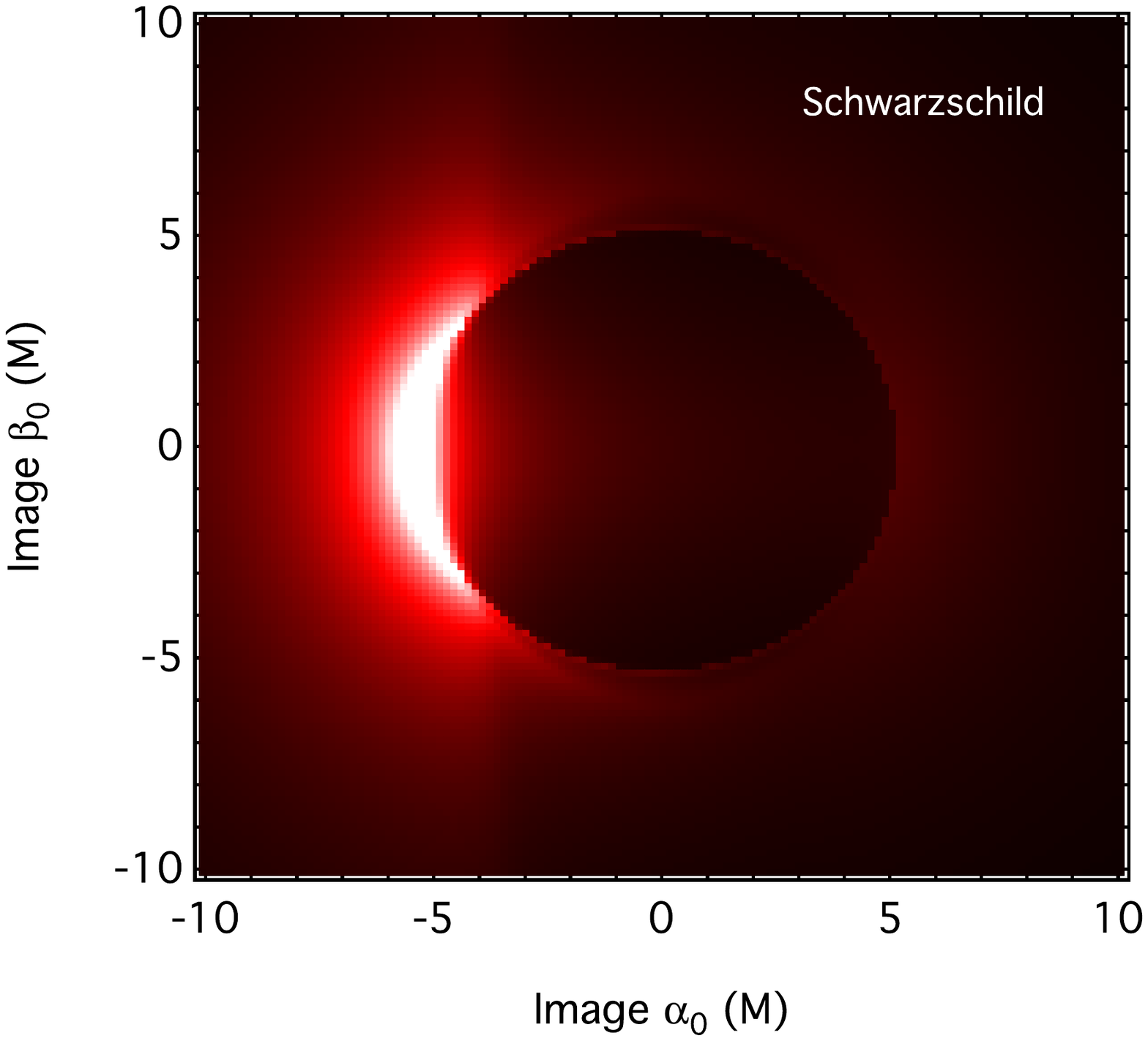}
  \includegraphics[scale=0.4, bb=3 7 510 461]{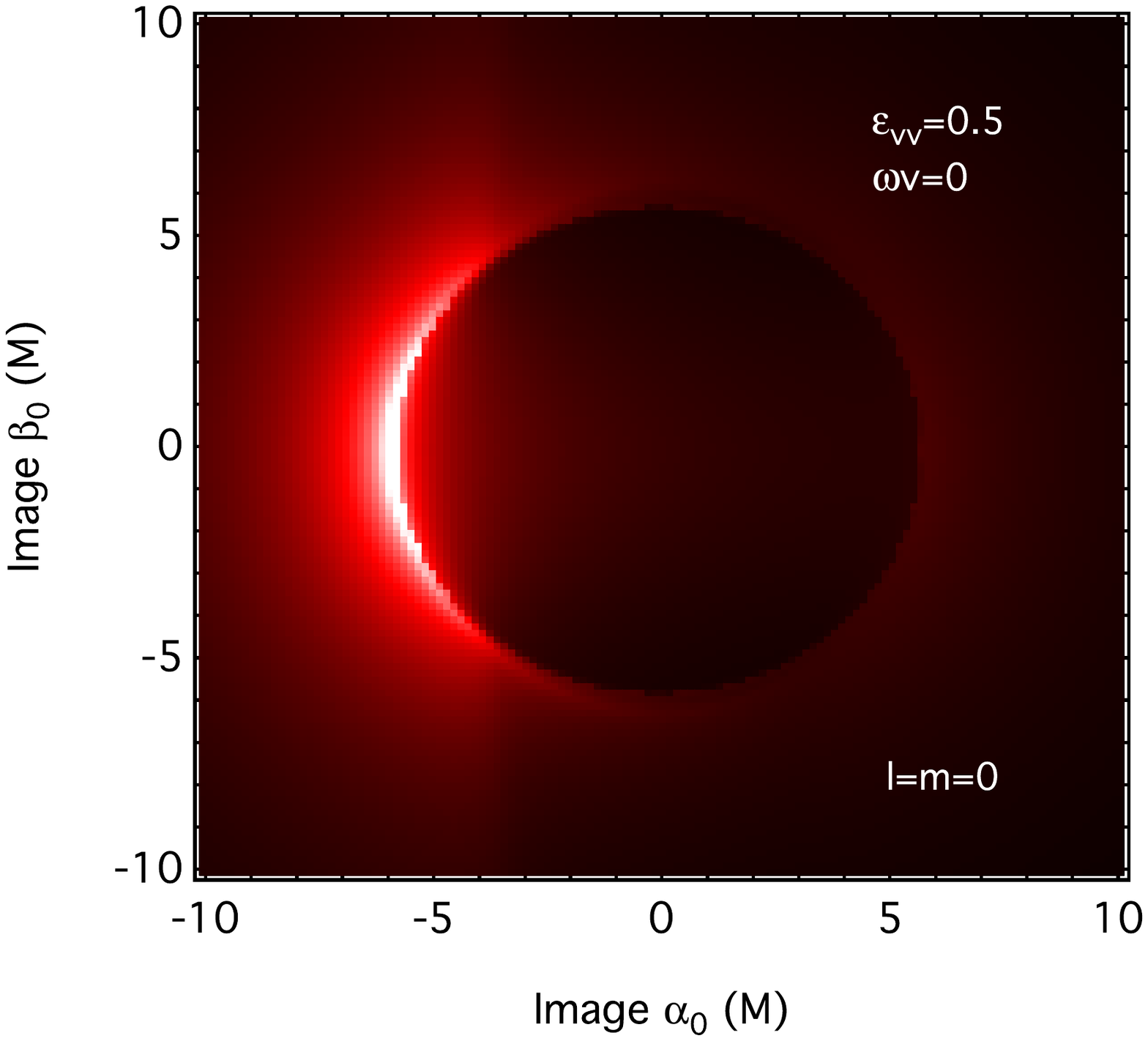}
  \includegraphics[scale=0.4, bb=3 7 510 461]{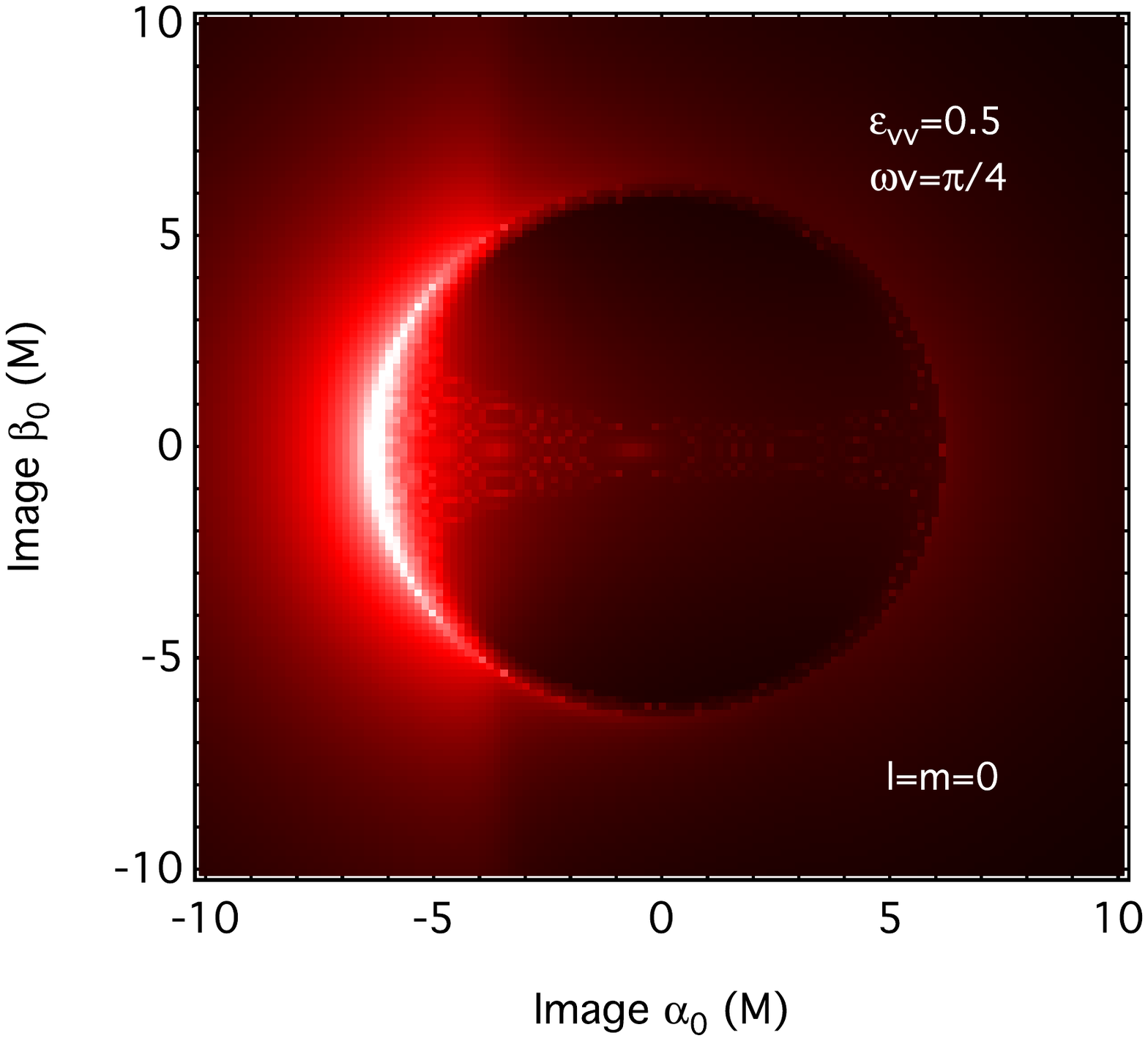}
  \includegraphics[scale=0.4, bb=3 7 510 461]{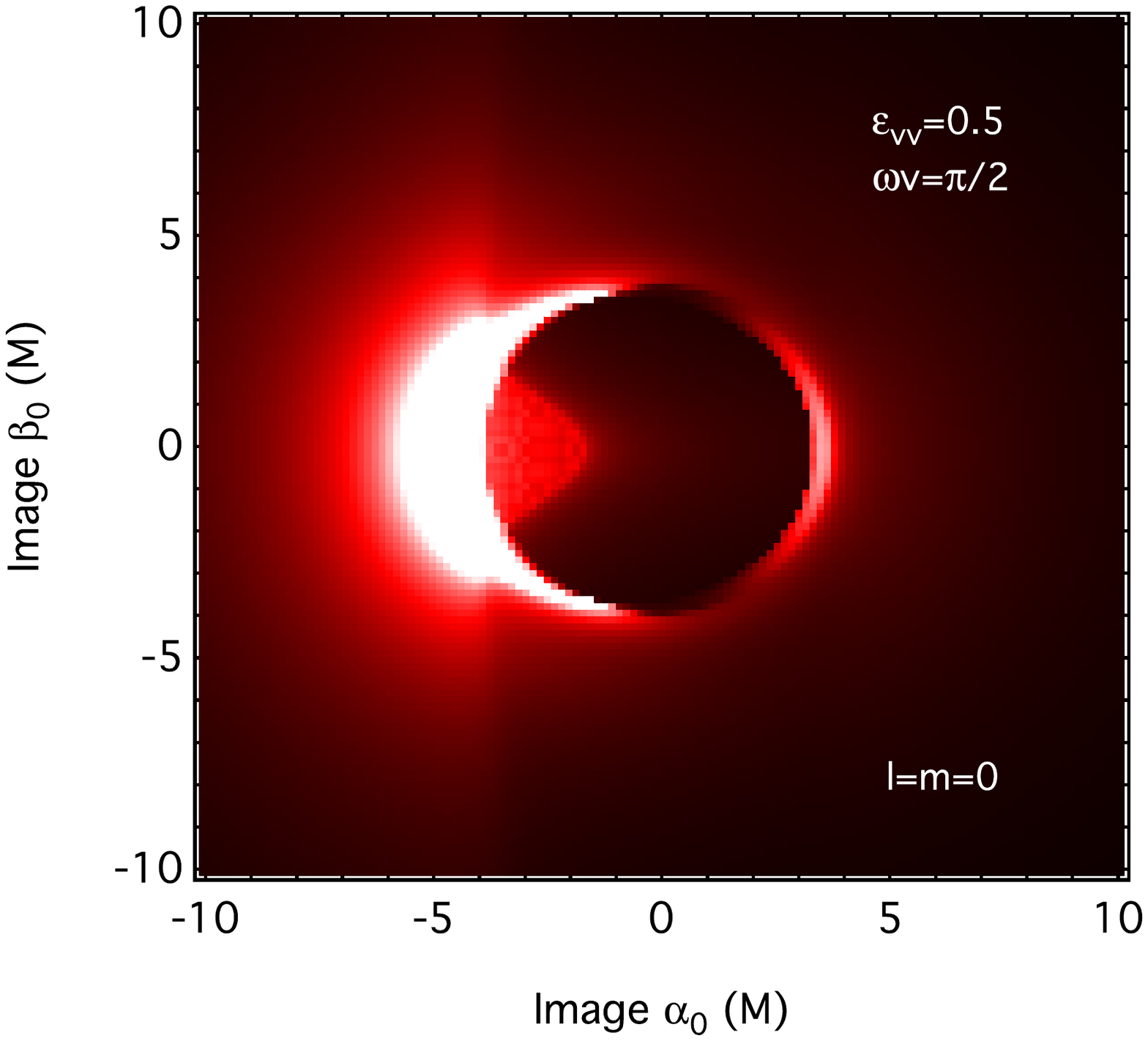}
  \caption{\footnotesize Same as Figure~\ref{fig:shadows} but with
    images from a simple plasma model for the emission from the inner
    accretion flow. The metric perurbations change in a time-dependent
    manner the image of the flow as well as the size, brightness, and
    width of the bright photon ring that surrounds the black-hole
    shadow.}
\label{fig:images}
\end{figure*}

\section{Images of Accretion Flows}

In this section, we explore the effect of the metric perturbations
discussed in \S3 on the expected images of accretion flows that will be
obtained by the Event Horizon Telescope. Following
Ref.~\cite{Broderick2014}, we consider the simplest possible plasma
model for the accretion flow that preserves the main characteristics
of images from more complicated GRMHD simulations (see, {\it e.g.\/},
\cite{Dexter2009, Dexter2010, Moscibrodzka2009, Moscibrodzka2014,
  Chan2015}).

The millimeter radiation observed from the main EHT targets is the
result of nearly optically thin synchrotron emission from thermal
electrons from a horizon-sized region in the accretion flow. The
monochromatic emissivity of synchrotron emission scales as $j\sim n_e
B^2$, where $n_e$ and $B$ are the electron density and magnetic field
of the region, respectively. The magnetic field is in near
equipartition with the plasma and, therefore, $B\sim n_e^{1/2}$. As a
result, the monochromatic emissivity scales as $j~\sim n_e^2$.

\begin{figure*}
  \includegraphics[scale=0.4, bb=3 7 510 461]{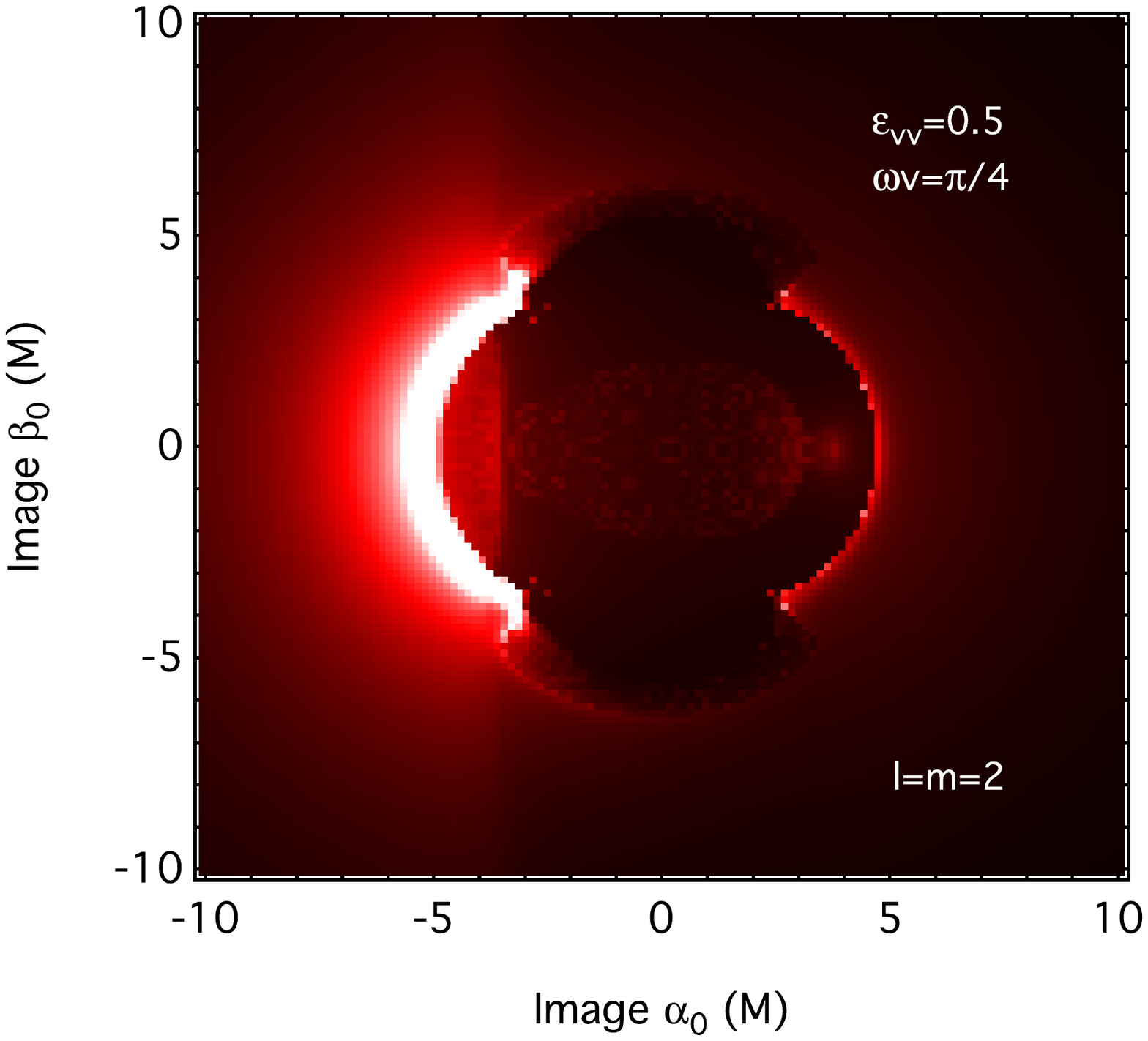}
  \includegraphics[scale=0.4, bb=3 7 510 461]{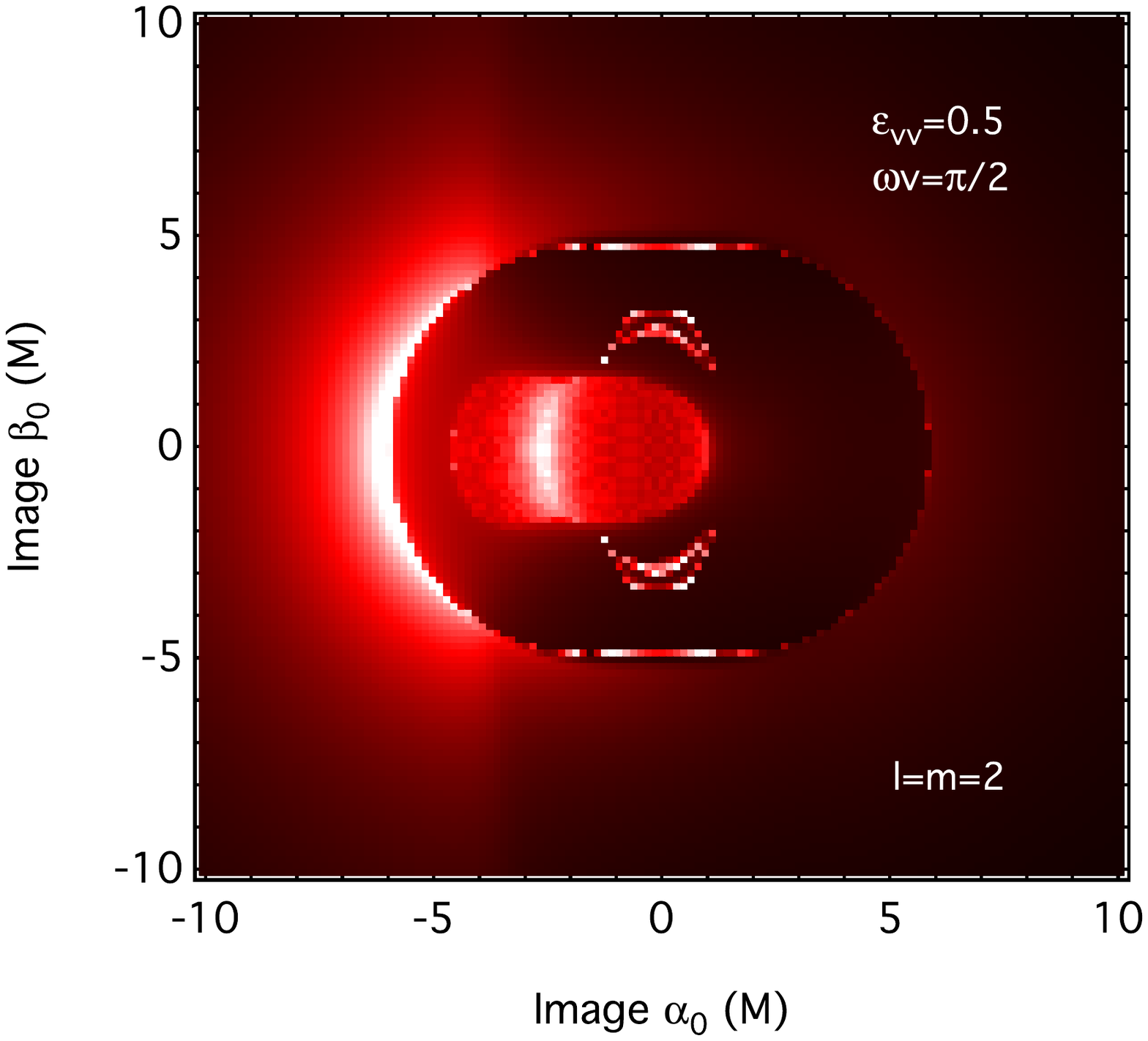}
  \caption{\footnotesize Same as Figure~\ref{fig:images} but
    for two phases of an $l=m=2$ perturbation mode.
    \label{fig:images_harm}}
\end{figure*}

\begin{figure*}
  \includegraphics[scale=0.4, bb=3 7 510 461]{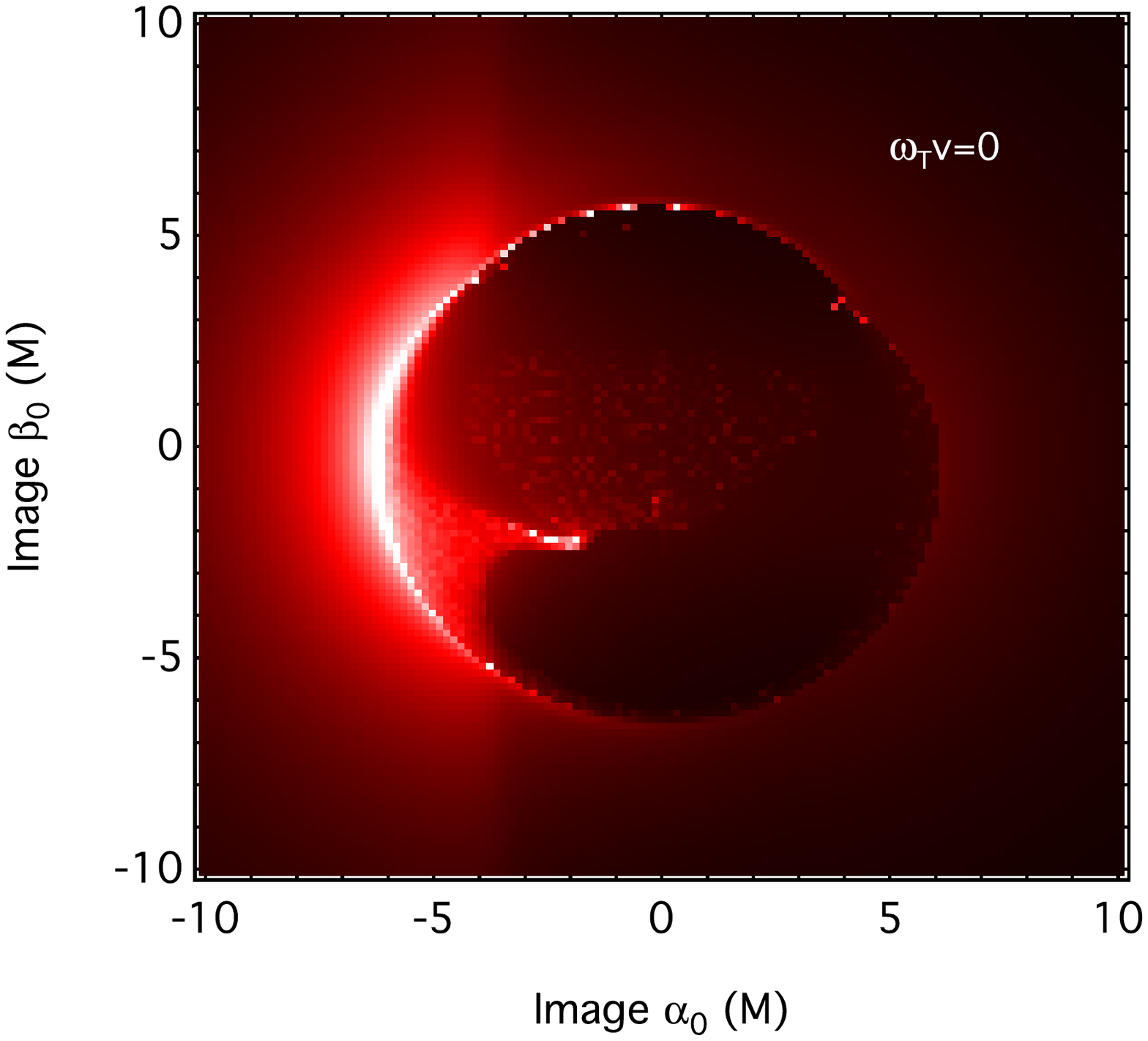}
  \includegraphics[scale=0.4, bb=3 7 510 461]{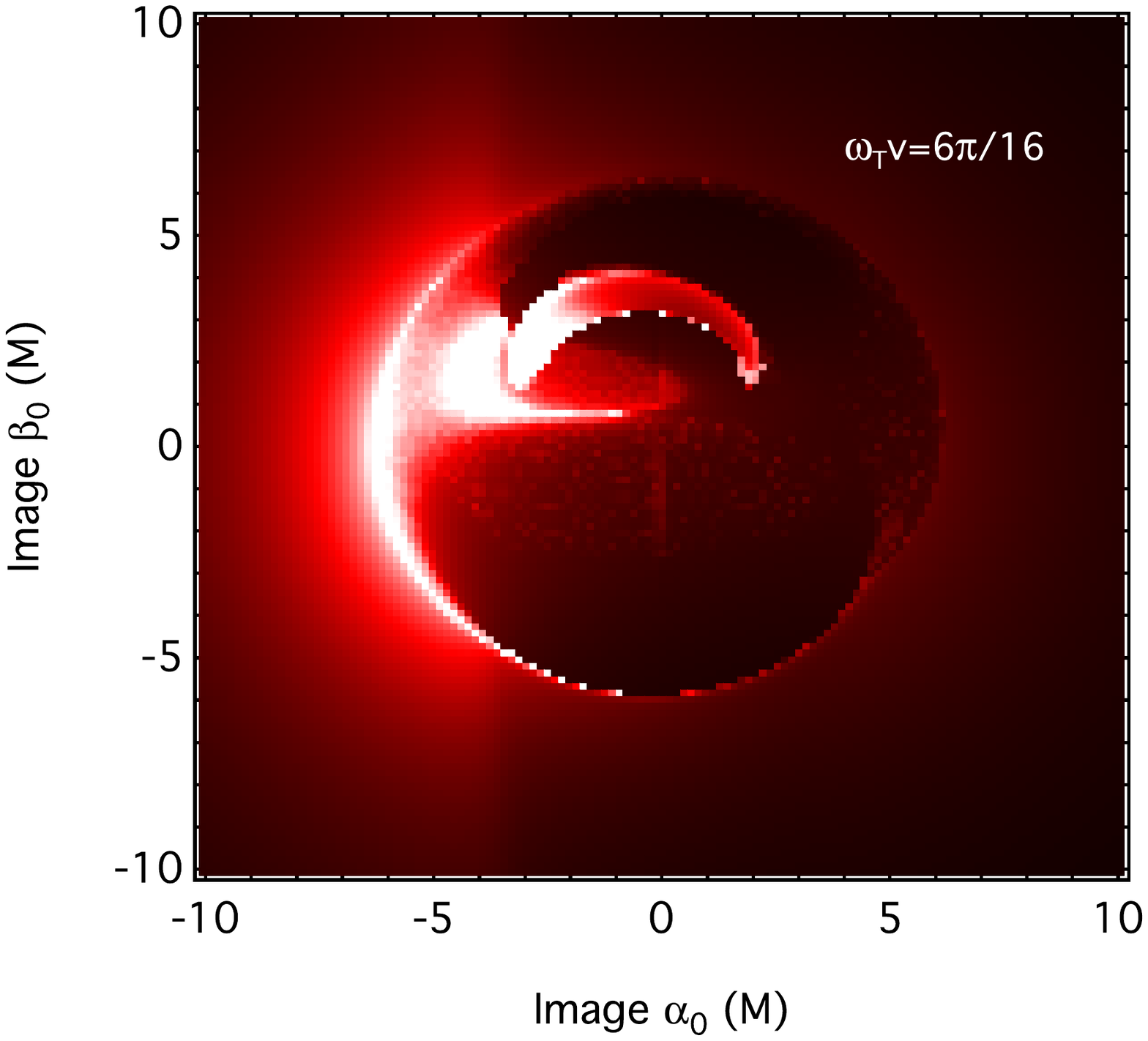}
  \includegraphics[scale=0.4, bb=3 7 510 461]{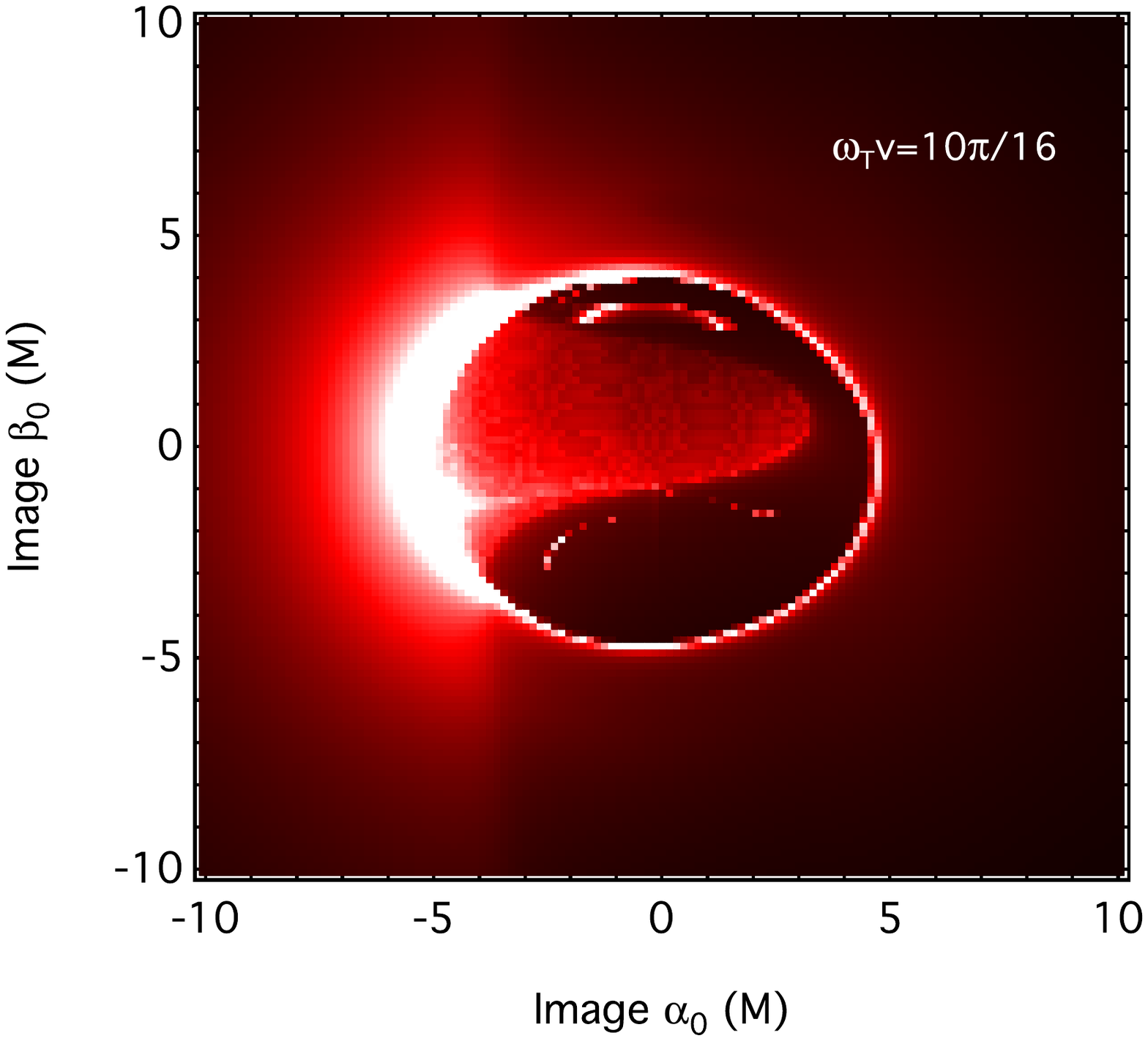}
  \includegraphics[scale=0.4, bb=3 7 510 461]{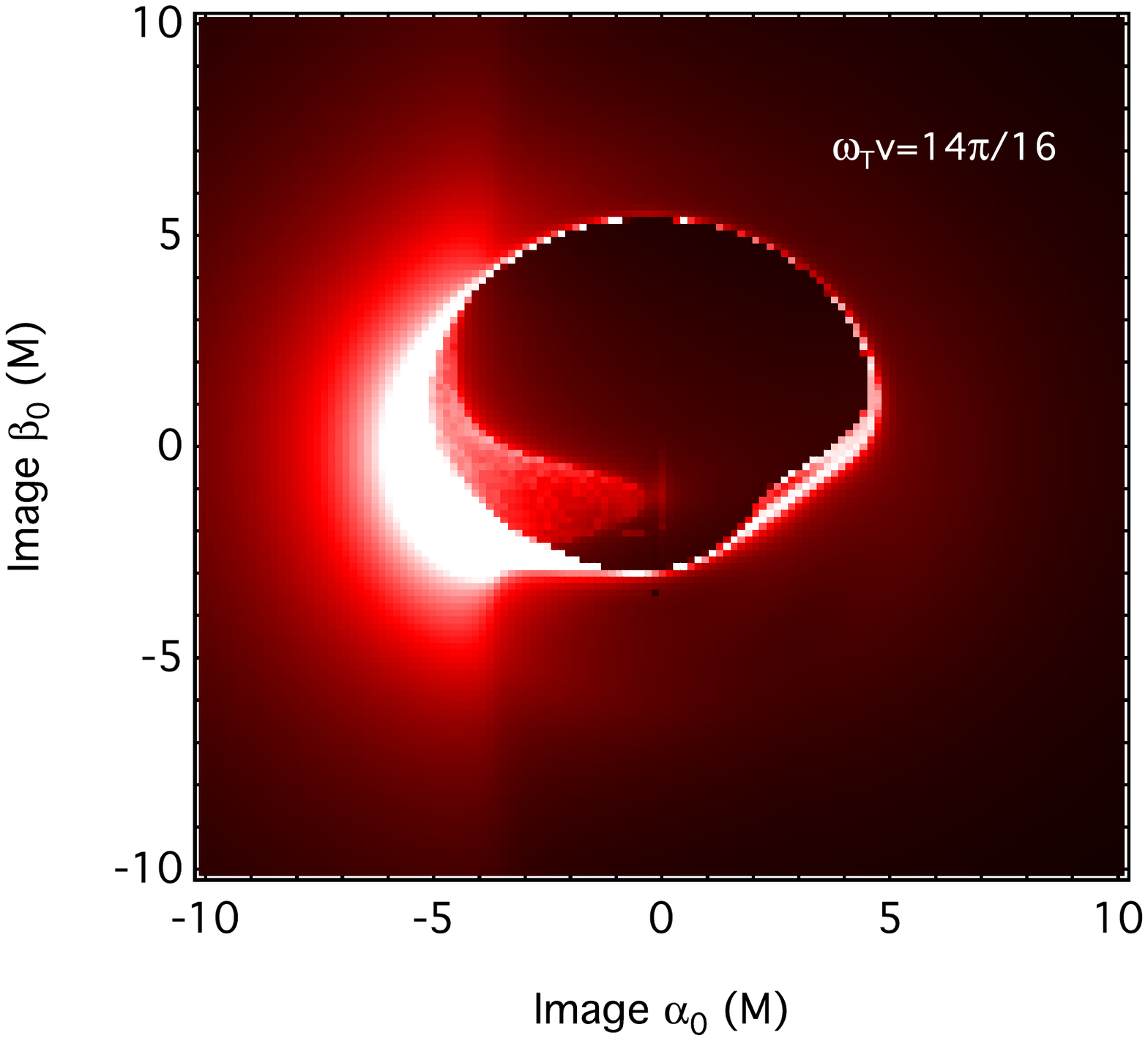}
  \caption{\footnotesize Four images of the accretion flow around a
    black hole at different instants with the spectrum of metric
    perturbations given in Table~\ref{table:modes}. The superposition
    of modes causes the shape and size of the black-hole shadow to be
    highly asymmetric and variable.
\label{fig:images_sup}}
\end{figure*}

\begin{figure*}
  \includegraphics[scale=0.4, bb=3 7 510 461]{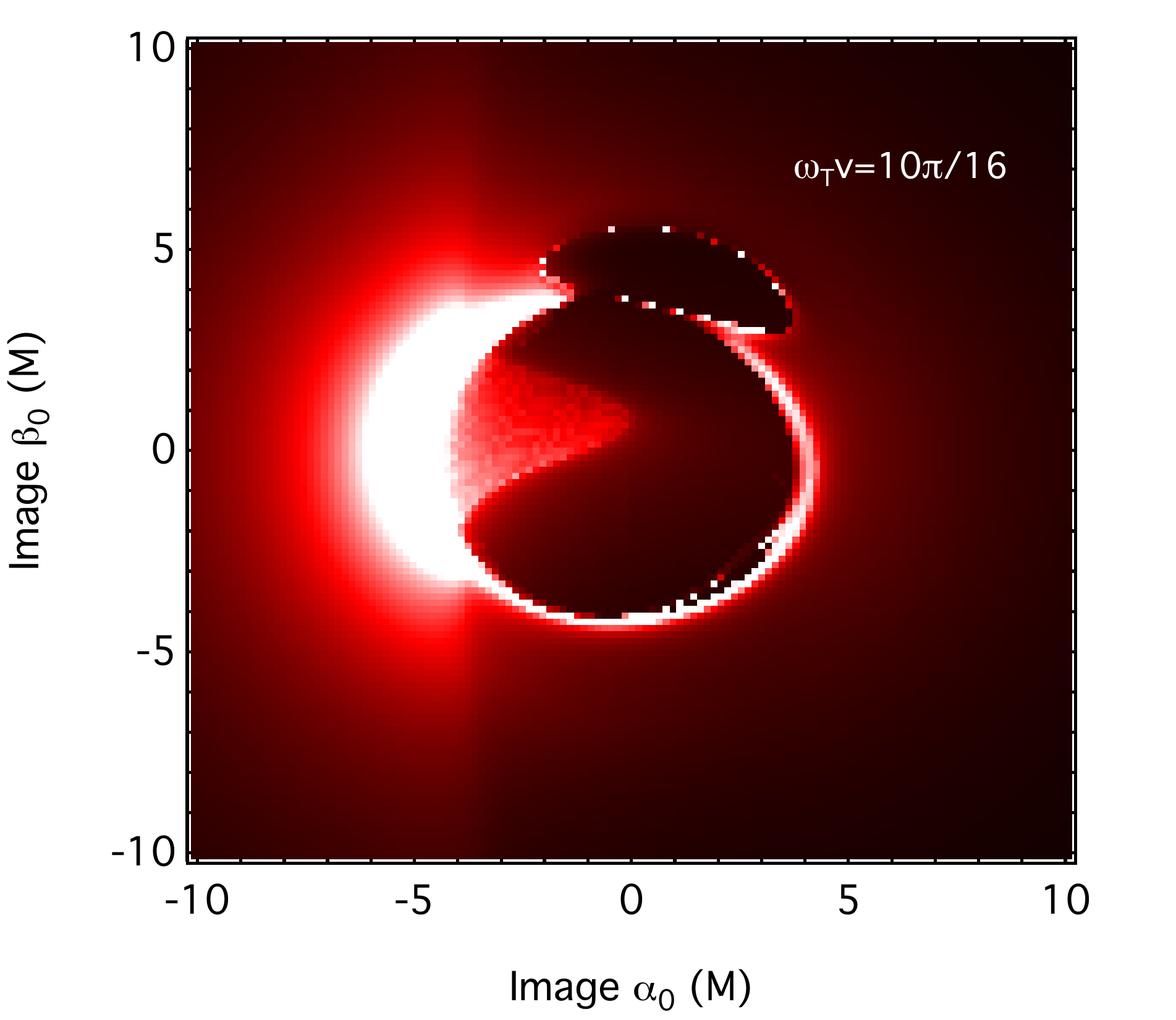}
  \includegraphics[scale=0.4, bb=3 7 510 461]{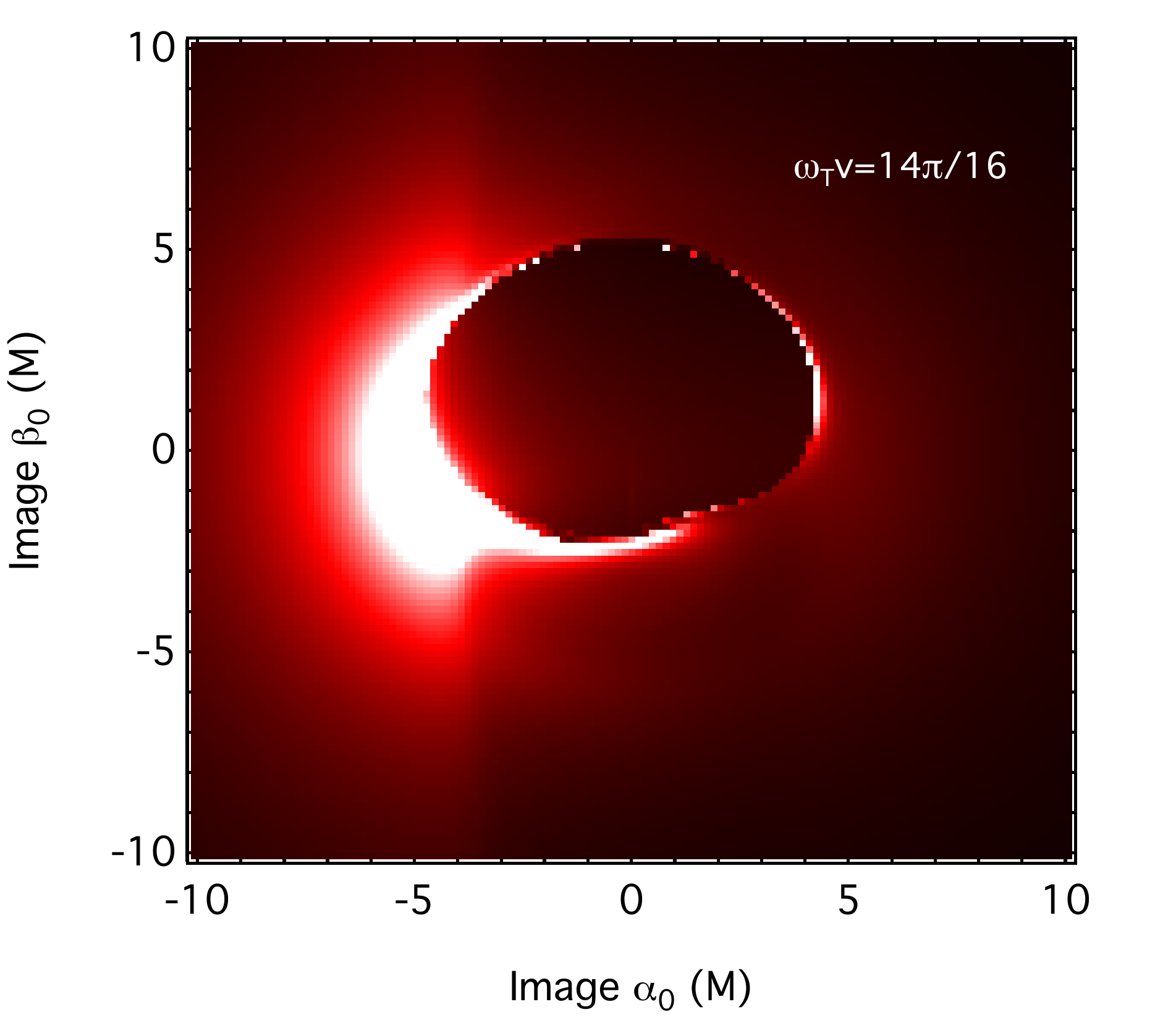}
  \caption{\footnotesize Same as Figure~\ref{fig:images_sup} but for
    all mode amplitudes reduced by a factor of three.
\label{fig:images_sup_epsilon}}
\end{figure*}

\begin{figure*}
  \includegraphics[scale=0.4, bb=3 7 510 461]{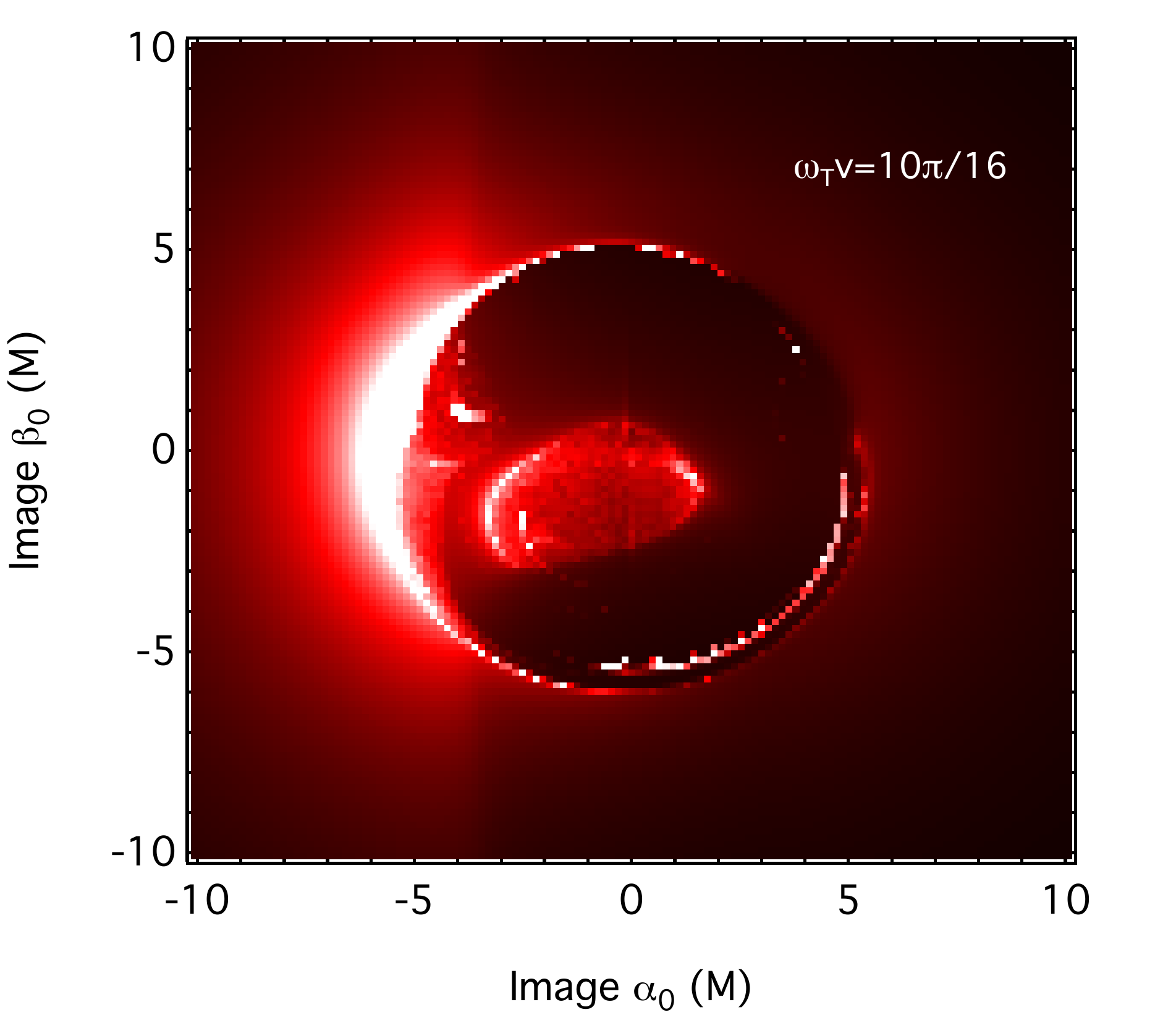}
  \includegraphics[scale=0.4, bb=3 7 510 461]{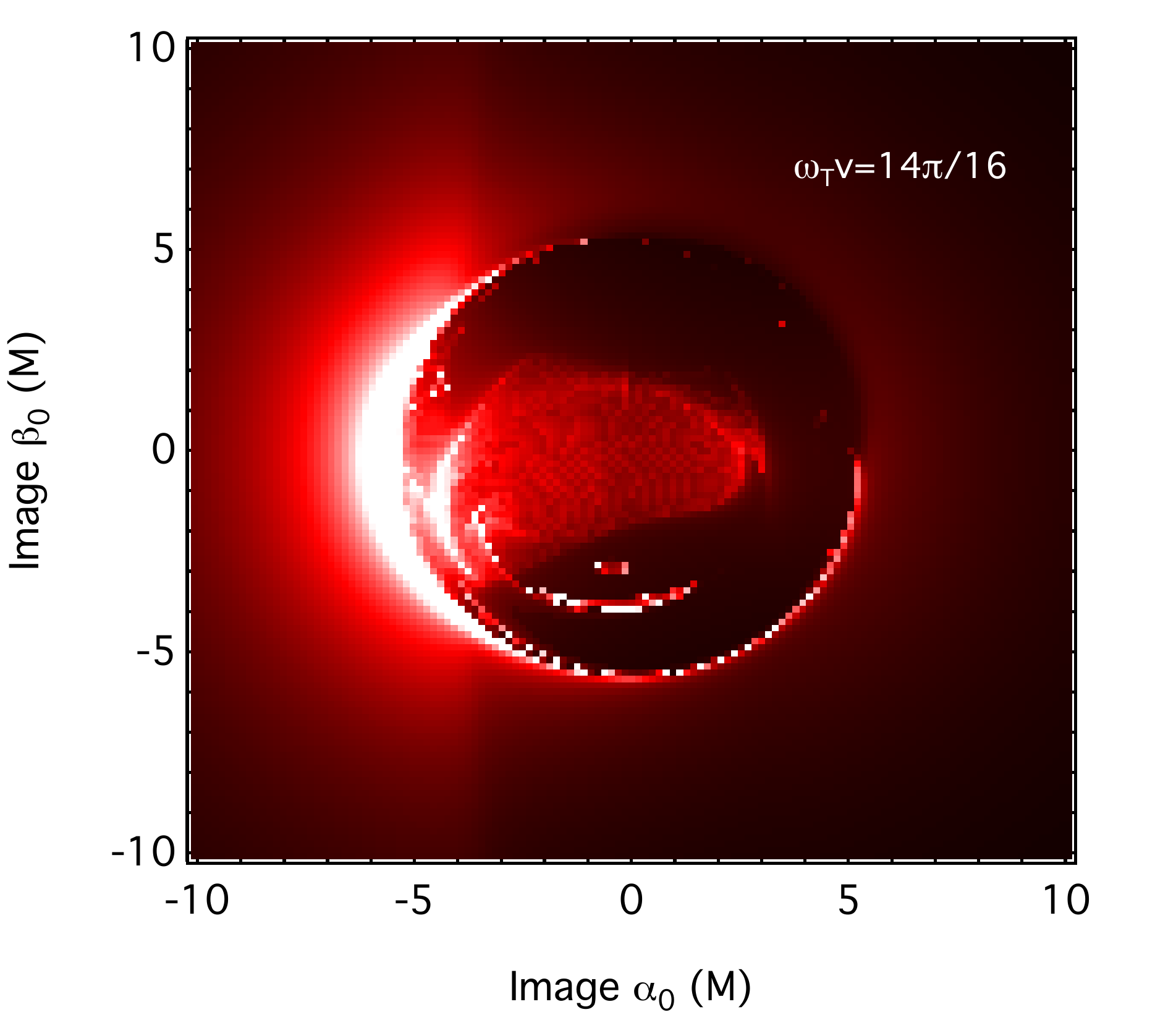}
  \caption{\footnotesize Same as Figure~\ref{fig:images_sup} but with all frequencies scaled 
  by artificially increasing to 
    $\omega_{\rm T}=1$.
\label{fig:images_sup_omega}}
\end{figure*}

Radiatively inefficient accretion models for the main EHT targets
suggest that the electron density in the inner flow region scales as
$n_e\sim r^{-1}$ (see \cite{Broderick2014} and references therein),
which we assume here. We also set the 4-velocity of the plasma outside
the location of the innermost stable circular orbit of the unperturbed
metric to the orbital velocity at the projected cylindrical radius
$\varpi=r\sin\theta$, {\it i.e.},
\begin{equation}
  \left(u^v, u^r, u^\theta, u^\phi\right)=
  \left(\sqrt{\frac{\varpi}{\varpi-3}},0,0,\frac{1}{\varpi\sqrt{\varpi-3}}\right)\;.
\end{equation}

 In the plunging region inside the innermost stable circular
  orbit, we set the plasma 4-velocity to the local free fall velocity
  of the unperturbed metric. We also set the local electron density
  such that the mass accretion rate in the plunging region is
  constant, i.e., such that
\begin{equation}
(n_{\rm e} u^\mu)_{;\mu}=0\;.
\end{equation}
In all the above, we are making the assumption that the plasma
properties are not affected by the metric perturbations. Even though
we make this assumption purely for simplicity, it is also justified by
the fact that most of the emission comes from radii $r>6$ at which the
metric perturbations are exponentially very small ({\it i.e.\/},
$\exp(-8)\simeq 3\times 10^{-4}$).

 The upper left panel of Figure~\ref{fig:images} shows the
  predicted image from this simplified model for an unperturbed
  Schwarzschild metric, as viewed by an observer at the equatorial
  plane of our coordinate system. The image is not spherically
  symmetric because of the cylindrical character of the plasma orbits
  and is dominated by the Doppler boosted crescent of emission from
  the plasma with velocities pointing towards the distant observer.
It also shows the outline of the black hole shadow and a faint photon
ring surrounding it. The remaining panels of the figure show the
images during different phases of a perturbation with the same
parameters as those shown in Figure~\ref{fig:trajectories}.

As discussed in \S4, the main effect of the metric perturbation is to
alter the location of the photon orbit and hence the size of the
shadow cast by the black hole on the surrounding emission. Moreover,
the radial wavelength of the perturbations causes trajectories with
different impact parameters to bundle up and this leads to pronounced
bright structures, such as the bright ring surrounding the shadow
(compare, {\it e.g.\/}, the lower right panels of
Figures~\ref{fig:trajectories} and \ref{fig:images}).

The presence of modes that are not spherically symmetric further
distorts the image of the accretion flow. Two example phases are shown
in Figure~\ref{fig:images_harm} for an $l=m=2$ mode with all the other
parameters being the same as for the $l=m=0$ mode shown in
Figure~\ref{fig:images}.

\begin{table*}
  \caption{\label{table:modes}Properties of sample spectrum of metric
  perturbations}
\begin{ruledtabular}
\begin{tabular}{cccccc}
  Mode & $lm$ & Amplitude & Phase$/2\pi$ & $\omega/\omega_{\rm T}$ &
  $k/\omega_{\rm T}$\\
  \hline
  $vv$ & 00 & 0.9 & 0.0 & 1 & $\pi/4$\\
  $vv$ & 21 & 1.0 & 0.2 & $\pi/4$ & $\sqrt{2}$ \\
  $rr$ & 00 & 1.1 & $\sqrt{2}$ & $\sqrt{2}$ & 1.1\\
  $rr$ & 11 & $\pi/3$ & 1.2 & 0.9 & 0.8\\
  $rv$ & 10 & $\pi/4$ & 0.3 & 1.1 & 0.9\\
  $rv$ & 20 & $\sqrt{2}$ & 0.7 & 0.8 & $\pi/3$ \\
  $\theta\theta$ & 10 & 1.2 & $\sqrt{5}/2$ & $\pi/3$ & $\sqrt{5}/2$ \\
  $r$ (odd) & 21 & $\sqrt{5}/2$ & 0.1 & 1.0 & $\pi/3$\\
  $v$ (odd) & 11 & 1.0 & 0.5 & 0.95 & 1.0
\end{tabular}
\end{ruledtabular}
\end{table*}

More realistic situations are, of course, expected to require us to
consider a superposition of oscillatory modes with different
amplitudes, frequencies, phases, and wavenumbers. This superposition
will break the symmetries as well as the purely periodic character of
the images shown in
Figures~\ref{fig:images}-\ref{fig:images_harm}. Moreover, it is
possible that the lack of symmetry might suppress the magnitude of the
effect on the images. Exploring the implications for the predicted
images from different spectra of perturbations is beyond the scope of
this initial study. However, in order to get a first look into the
effect of the superposition of different modes, we performed a
calculation with a small number of modes with properties summarized in
Table~\ref{table:modes}. The amplitudes, frequencies, wavelengths, and phases of
these modes were chosen rather arbitrarily, with the requirement that the
frequencies and wavelengths are non-commensurate.

Figure~\ref{fig:images_sup} shows the images calculated during four
instances in the evolution of the mode spectrum shown in
Table~\ref{table:modes}. The lack of symmetry in the images is
evident. Moreover, the superposition of modes does not suppress the
effect of individual modes on the images but rather exaggerates
it. Indeed, the black-hole shadow shape becomes highly asymmetric and
rapidly fluctuating and its characteristic ``size'' shows order unity
variability.

Figures~\ref{fig:images_sup_epsilon} and
  \ref{fig:images_sup_omega} show the effect of changing the overall
  amplitudes and the characteristic frequencies of the spectrum of
  perturbations. As expected from the discussion of the results
  presented in Figure~\ref{fig:opaq},  moderate reduction of the amplitude of
  perturbations does not introduce qualitative differences in the
  resulting images; the outline of the black-hole shadow is determined
  by the size and shape of the unstable photon orbits, which can be
  altered substantially even with moderately small metric perturbations. 
  (Fig.\ref{fig:opaq} suggests significant differences after an order-of-magnitude reduction.) 
  On the
  other hand, increasing the frequencies of the modes reduces the
  magnitude of the effect on the images, because the rapid
  oscillations of the various metric elements are averaged out during
  the time it takes for photons to cross distances comparable to the
  radius of the photon orbit.  

\begin{figure}
  \includegraphics[scale=0.4, bb=3 7 510 461]{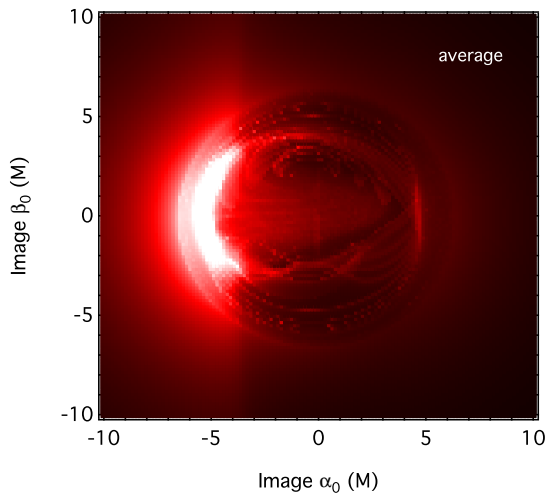}
  \caption{\footnotesize The average image for a black hole with the
    spectrum of metric perturbations given in Table~\ref{table:modes}
    calculated between the times $v=0$ and $v=32(2\pi/\omega_{\rm
      T})$. The net effect of averaging is a blurry black-hole shadow
    that is difficult to discern.
\label{fig:image_sup_ave}}
\end{figure}

Figure~\ref{fig:image_sup_ave} shows the average image of the
simulations shown in Figure~\ref{fig:images_sup} calculated between
the times $v=0$ and $v=32(2\pi/\omega_{\rm T})$. As expected, the
rapid variability of the images leads to a blurry black-hole shadow
that is difficult to discern.

\section{Implications for Event Horizon Telescope Observations}

In the last two sections, we showed that horizon-scale perturbations
in the spacetime of an astrophysical black hole introduce a
significant, time-dependent variation of the size and shape of the
shadow it casts on the surrounding emission, as well as bright
structures, such as rings of emission, that are also time dependent.
As discussed in \S2, we considered a characteristic frequency of these
oscillations that is of order $\omega_T=\left(8\pi M\right)^{-1}$ and
 which, therefore, depends on the mass of the black hole under
study.

There are two primary targets for the Event Horizon Telescope: the
black hole in the center of the Milky Way, Sgr~A$^*$, and the black
hole in the center of the M87 galaxy.  The former has a mass of $\sim
4.3\times 10^{6} M_\odot$~\cite{Ghez2008, Gillessen2009} while the
latter is $\sim 1500$ times more massive at $\sim 6.5\times 10^9
M_\odot$~\cite{Gebhardt2011}.  Even though it remains to be determined
whether a clearly visibile black-hole shadow is cast on their 1.3~mm
images, early EHT observations of both targets have demonstrated that
the image sizes are comparable to their horizons~\cite{Doeleman2008,
  Doeleman2012}.

Introducing physical units back to the dimensionless expressions, the
characteristic period of the metric perturbations becomes
\begin{eqnarray}
  P\simeq \frac{2\pi}{\omega_{\rm T}} = 
  &\simeq & 0.93 \left(\frac{M}{4.3\times 10^6 M_\odot}\right)~\mbox{hr}
  \nonumber\\
  &\simeq & 59 \left(\frac{M}{6.5\times 10^9 M_\odot}\right)~\mbox{d}\;,
\end{eqnarray}
where the first of the two estimates corresponds to Sgr~A$^*$ and the
second to M87. These are the timescales at which the black
  hole shadows will change, if quantum fluctuations such as the ones
  we consider here, are present.

The Event Horizon Telescope uses the rotation of the Earth to increase
the coverage of the interferometric $u-v$ plane and construct an
image. Each imaging observation will require scanning the $u-v$ plane
over a period of several hours.  In the case of Sgr~A$^*$, the
timescale of metric perturbations is comparable to or shorter than
each imaging scan and, therefore, individual snapshots of the effect
of metric perturbations on the images will not be generated. Moreover,
because of the particular interferometric methods used in the Event
Horizon Telescope (as in all mm VLBI experiments), the images
generated using multi-hour scans of the $u-v$ plane will not be simple
averages of the actual images, as the one shown in
Figure~\ref{fig:image_sup_ave}. This is because the Event Horizon
Telescope measures separately the magnitudes of the Fourier components
of the individual snapshots and the closure phases of the same Fourier
components along triangles of baselines, neither of which is the
result of liner operations on the images.

Different sources of variability in the black-hole image such as those
introduced by the turbulent accretion flow~\cite{Lu2016,
  Medeiros2016}, by refractive scattering in the interstellar
medium~\cite{Johnson2015a}, as well as by the metric perturbations we
discuss here, will imprint different signatures on the interferometric
visibilities.  Non-imaging techniques such as those discussed in
Ref.~\cite{Doeleman2009} will need to be employed to distinguish and
identify the signatures of metric perturbations from other sources of
variability.  Distinguishing characteristics  of the metric perturbations we consider
include the fact that
  they decay very quickly away from the horizon,
  causing them to alter primarily the outline of the black-hole shadow
  without introducing significant variability in the bulk of the
  emission, which originates outside the radius of the innermost
  stable circular orbit. Moreover, only metric perturbations are
  achromatic, {\it i.e.}, they affect photons of all
  wavelengths in the same way. In the near future, the Event Horizon
  Telescope will operate at two different wavelengths (1.3~mm and
  0.8~mm). Comparison of the image variability at these two
  wavelengths will be  important in separating the effects of
  perturbations in the geometry of spacetime from those introduced by
  variability in the plasma.

In the case of M87, the assumed timescale of metric perturbations is
longer than a single imaging scan and, therefore, images constructed
in different observing epochs (days) will correspond to different
snapshots of the perturbations. Moreover, the lack of appreciable
image blurring due to scattering in the interstellar medium towards
M87 will allow for measuring the size and shape of the black hole
shadow in that source without a need for additional corrections. Both
these properties make the black hole in M87 the optimal candidate for
searching for black hole quantum structure with the Event Horizon
Telescope.

 If evidence of perturbations like those considered here is found in EHT observations, there 
is no apparent explanation within classical GR.  Thus, 
this would strongly  indicate phenomena well-outside a general-relativistic description, with a prime 
candidate being quantum black hole structure associated with the need to make black holes consistent with quantum mechanics.

\section{Acknowledgements}

We wish to thank S.\ Britzen, D.\ Marolf, and F.\ \"Ozel, for helpful
discussions.  The work of SG was supported in part by the U.S.\ DOE
under Contract No.\ DE-SC0011702 and by Foundational Questions
Institute (fqxi.org) grant FQXi-RFP-1507. The work of DP was supported
in part by NASA TCAN award NNX14AB48G and by NSF grant AST~1312034.

\bibliographystyle{utphys}

\bibliography{eht}

\end{document}